\documentclass{jfm}
\usepackage{natbib}
\usepackage{graphicx}
\usepackage{epstopdf, epsfig}
\usepackage{caption}
\usepackage{subcaption}
\usepackage{gensymb}
\usepackage{color}
\usepackage{amsmath,mathtools}
\usepackage{amsfonts}
\usepackage{amssymb}
\usepackage{amsmath} 
\usepackage{upgreek}
\usepackage{mathrsfs}
\usepackage{float}
\usepackage{hyperref}
\usepackage[normalem]{ulem}

\newcommand{\be}{\begin{equation}}
\newcommand{\ee}{\end{equation}}

\newcommand{\aproxsol}[1]{\mathtt{#1}}
\newcommand{\edit}[1]{\textcolor{black}{#1}}

\newcommand{\We}{W\kern-0.15em e\,}
\newcommand{\Bo}{B\kern-0.05em o\,}
\newcommand{\Oh}{O\kern-0.05em h\,}
\newcommand{\R}{R\kern-0.15em e\,}

\newdimen\figrasterwd
\figrasterwd\textwidth

\shorttitle{Drop rebound at low Weber number}
\shortauthor{C.T. Gabbard and others}

\title{Drop rebound at low Weber number}

\author{Chase T. Gabbard$^{1, \corresp{\email{chase\_gabbard@brown.edu}}}$, Elvis A. Ag\"uero$^{1}$, Radu Cimpeanu$^{2}$, Katharina Kuehr$^{3}$, Eli Silver$^{1}$, Jack-William Barotta$^{1}$, Carlos A. Galeano-Rios$^{3}$, \and Daniel M. Harris$^{1, \corresp{\email{daniel\_harris3@brown.edu}}}$}

\affiliation{$^{1}$School of Engineering, Brown University, Providence, RI 02912, USA  \\ $^{2}$Mathematics Institute, University of Warwick, Coventry CV4 7AL, UK \\ $^{3}$College of Computational Sciences, Minerva University, San Francisco, CA 94103, USA}

\begin{document}

\maketitle

\begin{abstract}
We study the rebound of drops impacting non-wetting substrates at low Weber number $\We$ through experiment, direct numerical simulation, and reduced-order modeling. Submillimeter-sized drops are normally impacted onto glass slides coated with a thin viscous film that allows them to rebound without contact line formation. Experiments are performed with various drop viscosities, sizes, and impact velocities, and we directly measure metrics pertinent to spreading, retraction, and rebound using high-speed imaging. We complement experiments with direct numerical simulation and a fully predictive reduced-order model that applies natural geometric and kinematic constraints to simulate the drop shape and dynamics using a spectral method. At low $\We$, drop rebound is characterized by a weaker dependence of the coefficient of restitution on $\We$ than in the more commonly studied high-$\We$ regime, with nearly $\We$-independent rebound in the inertio-capillary limit, and an increasing contact time as $\We$ decreases. Drops with higher viscosity or size interact with the substrate longer, have a lower coefficient of restitution, and stop bouncing sooner, in good quantitative agreement with our reduced-order model. \edit{In the inertio-capillary limit, low $\We$ rebound has nearly symmetric spreading and retraction phases and a coefficient of restitution near unity. Increasing $\We$ or viscosity breaks this symmetry, coinciding with a drop in the coefficient of restitution and an increased dependence on $\We$.} Lastly, the maximum drop deformation and spreading are related through energy arguments, providing a comprehensive framework for drop impact and rebound at low $\We$. 
\end{abstract}

\begin{keywords}
drops, capillary flows
\end{keywords}

\section{Introduction}\label{sec:introduction}

Drop impacts are common in everyday life, industry, and nature. Examples include drops splashing on a kitchen sink, precision inkjet printing \citep{lohse2022fundamental}, and raindrops striking leaves and animal fur \citep{roth2022rain}. Since the seminal studies of Worthington \citep{worthington1877xxviii,worthington1908study} and Edgerton's iconic images \citep{edgerton1954flash}, researchers have strived to uncover the physics of impact phenomena, which include drop deposition, splashing, and rebound \citep{rein1993phenomena,yarin2006drop}. Early studies primarily focused on splashing, but interest in drop rebound has surged, inspired in part by raindrops bouncing off superhydrophobic leaves \citep{chen2011comparative} and driven by advances in surface preparation techniques that can render surfaces hydrophobic and enable rebound \citep{richard2000bouncing,okumura2003water}. Despite the impact of these innovations on numerous everyday items, such as windshields and eyewear, our understanding of the physics governing rebound is mostly limited to millimeter-sized drops striking surfaces at moderate and high speeds. This paucity motivates a detailed study of low speed rebound of submillimetric drops, which are typical in commercial spray nozzles \citep{makhnenko2021review,chen2022droplet}, naturally emerge from ligament breakup \citep{villermaux2020fragmentation}, and can result from splashing of millimetric drops \citep{yarin2006drop}. Recent attempts to fill this knowledge gap have been afflicted with contact line friction, which increasingly alters the rebound physics as the impact velocity decreases \citep{wang2022successive,thenarianto2023energy}. To address this knowledge gap, we perform experiments on a perfectly non-wetting substrate---a thin viscous film---using submillimeter-sized drops and identify the relevant physics governing rebound using reduced-order modeling, direct numerical simulation, and energetic arguments.

Over a century ago, Lord Rayleigh observed drops rebounding upon impact \citep{rayleigh1879capillary} and attributed this phenomenon to an intervening air layer separating the drops and preventing coalescence \citep{rayleigh1899xxxvi}. Since then, experiments conducted under modified atmospheric conditions and corresponding theoretical studies have confirmed his hypothesis \citep{willis2000experiments,bach2004coalescence,chubynsky2020bouncing}. Meanwhile, advancements in experimental and computational techniques have enabled measurement of the minimum thickness of the air layer---on the order of 100 nm---below which van der Waals forces bridge the liquid-air interfaces and capillary forces drive coalescence \citep{couder2005bouncing,de2012dynamics,kolinski2012skating,kavehpour2015coalescence,sprittles2024gas,lewin2024collision}. Thus, drops only rebound without contact from smooth surfaces. Although there are limited solid surfaces smooth enough to promote contactless rebound, liquid-gas interfaces are atomically smooth and ideal for rebound. For example, recent results for the rebound of a drop on a bath of the same fluid showed that the deformation of the drop and bath are dynamically coupled \citep{alventosa2023inertio,Phillips_Milewski_2024}, and that vibrating the bath can lead to long-lasting bouncing and wave-propelled drop motion \citep{couder2005walking,molavcek2013drops}. This dynamic coupling is further evidenced by the complex bubble patterns observed when an impacting drop merges with a bath \citep{saylor2012experimental}. If the bath height is reduced to a thin film, rebound can become substrate-independent \citep{hao2015superhydrophobic,sanjay2023drop}. Although drops can rebound from heated substrates \citep{biance2006elasticity}, dry ice \citep{li2024rebound}, and when the drop is coated with hydrophobic powder \citep{aussillous2006properties}, these methods inject energy into the system or modify the drop composition, potentially altering the rebound dynamics and obfuscating the fundamental physics.

\begin{figure}
  \centerline{\includegraphics[width=135mm]{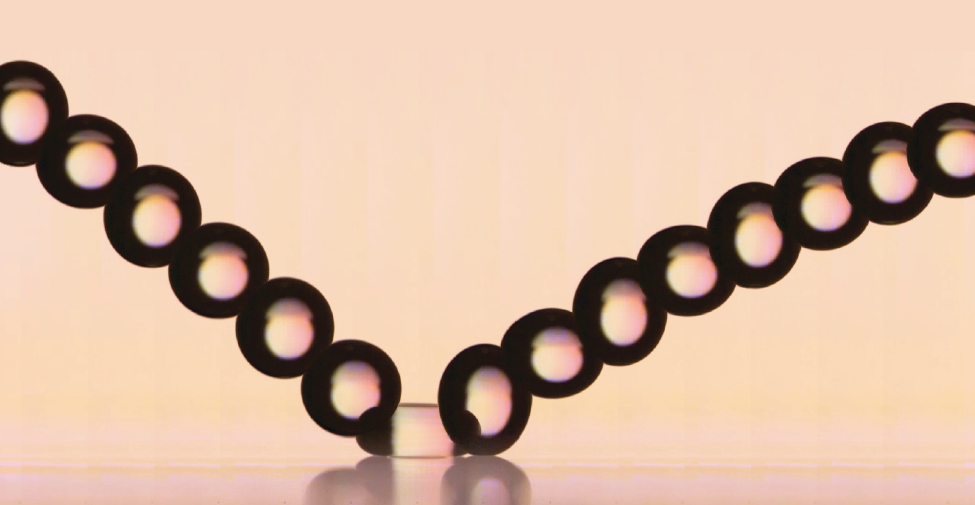}}
  \caption{A drop of silicone oil ($R = 0.31$ mm) rebounding from a glass slide coated with a thin, viscous film ($\Gamma / \Oh_{f}^{1/3} \approx 10^{-2}$). Here, $\We=1.03$, $\Oh=0.02$, and $\Bo=0.04$. Subsequent images are separated by 1.44 ms and shifted 0.4 mm to the right.}
\label{fig:Fig1}
\end{figure}

Rebound dynamics are governed by a competition of inertial, viscous, capillary, and gravitational forces. A common set of governing dimensionless parameters describing the problem are
\begin{equation}\label{eqn:dimensionlessparameters}
\begin{array}{ccc}
\quad \We = \frac{\rho \, V^2 \, R}{\sigma}, & \quad
\Bo = \frac{\rho \,g \, R^2}{\sigma}, & \quad
\Oh = \frac{\mu}{\sqrt{\rho \,\sigma \, R}}. \quad
\end{array}
\end{equation}
The Weber number $\We$ compares inertial and surface tension forces, Bond number $\Bo$ compares gravitational and surface tension forces, and Ohnesorge number $\Oh$ is the ratio of the inertio-capillary to \edit{the viscous} timescale. Here, $R$ is the undeformed drop radius, $V$ is the impact velocity, $\rho$ is the liquid density, $\mu$ is the dynamic viscosity, $\sigma$ is the surface tension coefficient, and $g$ is the gravitational acceleration. Although early studies emphasized $\We$ as the primary parameter controlling rebound \citep{foote1975water,anders1993velocity,richard2000bouncing,richard2002contact,okumura2003water}, recent work has shown that increasing $\Bo$ and $\Oh$ can suppress rebound for $\We>1$ \citep{jha2020viscous,sanjay2023does}. For $\We<1$, the role of $\Bo$ and $\Oh$ is less explored, and attempts to isolate the role of $\We$ on rebound have been confounded by the increasing influence of contact line friction as $\We$ decreases \citep{wang2022successive,thenarianto2023energy}. Impacting a drop on a substrate coated with a viscous film avoids contact line formation, as discussed by \citet{gilet2008dynamics} and \citet{gilet2012droplets}. This technique produces substrate-independent rebound \edit{when the film is sufficiently thin and viscous} \citep{hao2015superhydrophobic}. \edit{To determine the criteria where drop rebound is substrate-independent,} \citet{sanjay2023drop} analyzed the competing influences of the drop Ohnesorge number $\Oh$ and film Ohnesorge number $\Oh_{f}=\mu_{f} / \left( \rho \sigma R \right)^{1/2}$, and found that substrate-independent rebound occurs for $\Oh_{f} \gtrsim 0.1$ when $\Gamma < \textrm{B} \, \Oh_{f}^{1/3}$, where $\Gamma = h_{f}/R$ and $\textrm{B} = \textrm{B} \left( \Oh \right)$. Since $\textrm{B} \approx 0.1$ when $\Oh \approx 0.1$, and $\textrm{B}$ increases with decreasing $\Oh$, a conservative condition for substrate-independent bouncing is $\Gamma/\Oh_{f}^{1/3} < 0.1$. Notably, this result is derived for $\We >1$ and becomes less restrictive for $\We < 1$. By following these guidelines, a rigid substrate can be prepared that allows a drop to rebound without contact line formation or dynamic influence from the viscous film. An example of such a rebound is imaged in figure~\ref{fig:Fig1}. 

When a drop impacts a rigid substrate, it first spreads until reaching a maximum equatorial radius $R_m$ and maximum contact radius $R_c$. When $\We \gg 1 $, the dimensionless maximum equatorial radius $\beta = R_{m}/R$ and maximum contact radius $\beta_{c}= R_{c}/R$ are approximately equal, and two competing scaling laws have been proposed to describe their evolution with $\We$. \citet{clanet2004maximal} proposed $\beta \sim \We^{1/4}$ based on momentum conservation and assuming the drop deforms into a cylinder with negligible height during impact. Alternatively, the conservation of energy between the kinetic energy of the drop at impact and the surface energy at the maximum deformation of the drop predicts $\beta \sim \We^{1/2}$ \citep{chandra1991collision,bennett1993splat,okumura2003water}. Although additional arguments and empirical evidence favor the scaling $\We^{1/2}$ \citep{eggers2010drop,villermaux2011drop,laan2014maximum,lee2016universal,josserand2016drop}, extensive experimental results support the scaling $\We^{1/4}$ over a wide range of $\We$ \citep{bartolo2005retraction,biance2006elasticity,tsai2011microscopic,garcia2020inclined,zhan2021horizontal,li2024rebound}. When viscous forces are important, these scalings break down \citep{rein1993phenomena}, and describing the crossover from the capillary to viscous regime requires consideration of finite $\Oh$ effects (or $Re=\sqrt{\We}/\Oh=\rho V R / \mu$) \citep{roisman2009inertia,eggers2010drop,laan2014maximum,lee2016universal,sanjay2024unifying}. For $\We<1$, the deformed drop resembles an oblate ellipsoid rather than a wide cylinder and $\beta \not\approx \beta_{c}$. In this case, \citet{richard2000bouncing} predicted $\beta -1 = \left(5/48\right)^{1/2} \, \We^{1/2}$ by assuming the initial kinetic energy of the drop was converted to surface energy but did not test their prediction across a range of $\We$. Recently, efforts to describe and relate $\beta$ and $\beta_{c}$ during low $\We$ rebound have used drops with non-negligible $\Bo$ and relied on fits to empirical relations \citep{liu2023maximum,li2024rebound}. Thus, a comprehensive and validated understanding of spreading for low $\We$ is still lacking.

Following spreading, surface tension drives the drop to retract and can lead to rebound. When $\Oh$ is negligible, the maximum retraction rate is independent of impact velocity, and the retraction time $t_{r} \sim t_{\sigma}$, where $t_{\sigma}=\left( \rho R^{3} / \sigma \right)^{1/2}$ is the intertio-capillary time \citep{bartolo2005retraction}. This result is valid for $\We>1$, and since inertia also governs the spreading time $t_s$, the interaction time between the substrate and drop $t_{c}=t_{s}+t_{r}$, referred to as the contact time henceforth, scales as $t_c \sim A \, t_{\sigma}$ \citep{richard2002contact}, where $A=f\left(\Oh\right)$ \citep{jha2020viscous}. Alternatively, when $\We<1$, the non-dimensional contact time $t_c/t_{\sigma}$ increases as $\We$ decreases \citep{foote1975water,okumura2003water}. This transition in $\We$ dependence signals a shift in the underlying physics. This shift also alters the $\We$ dependence of the coefficient of restitution $\varepsilon$, a measure of the elasticity of the rebound. For $\We>1$, \citet{anders1993velocity} noted that $\varepsilon$ is $\We$ dependent and \citet{biance2006elasticity} showed that $\varepsilon \sim We^{-1/2}$, which has been verified on various rigid substrates \citep{wang2022successive,thenarianto2023energy}. Additionally, \citet{jha2020viscous} showed that increasing $\Oh$ decreases the coefficient of restitution $\varepsilon$. For $\We<1$, \citet{gilet2012droplets} suggested $\We$-independent rebound such that $\varepsilon = f \left( \Oh \right)$. Recent efforts to study $\varepsilon$ for low $\We$ rebounds observed a peak in $\varepsilon$ as $\We$ decreases, followed by a decrease to $\varepsilon = 0$ (no rebound) at finite $\We$ \citep{wang2022successive,thenarianto2023energy}. The suggested functional form of this decrease $\varepsilon \sim \We^{1/2}$ is attributed to contact line friction, highlighting the need for contact line-free conditions to determine the $\varepsilon$–$\We$ relationship at low $\We$.

These unique trends in the rebound metrics at low $\We$ have inspired several models. \citet{okumura2003water} derived a simple model for the contact time $t_{c}$ and maximum equatorial deformation $\beta$ by treating the drop as a spring-mass system, with spring constant scaling with the surface tension coefficient. \citet{jha2020viscous} expanded this model to include damping to describe the rebound dynamics of viscous drops. \citet{molavcek2012quasi} proposed a quasi-static model for impacts on non-wetting surfaces, approximating drop shapes as static during impact, giving analytical predictions for $t_c$ and $\varepsilon$ in various limits. Concurrently, \citet{chevy2012liquid} derived a quasi-static model for the droplet deformation, highlighting the non-linear elastic behavior of weakly deformed drops. More recently, \citet{galeano2017non} introduced a kinematic match (KM) method to derive a fully predictive model for the rebound of a rigid hydrophobic sphere on a liquid bath, which has been extensively validated through experiments and direct numerical simulation \citep{galeano2017non,galeano2019quasi,galeano2021capillary}. The KM method imposes natural kinematic and geometric constraints on the motion of interacting surfaces, ensuring continuity at the contact interface. This approach is particularly well-suited for rebound problems as it provides a robust framework to capture the interplay between fluid dynamics and interfaces, making it highly versatile for a wide range of impact scenarios in mechanics. For example, \citet{aguero2022impact} applied the KM method to non-Hertzian impact of a solid sphere on an elastic membrane and observed good agreement with experiment. These successful applications of the KM method and its ability to efficiently simulate rebound phenomena with minimal assumptions motivate its extension to drop rebound on rigid substrates.

Herein, we present experiments and direct numerical simulation results of drop rebound at low $\We$, and develop a robust reduced-order model using the KM method. The experimental apparatus, testing protocol, and image processing technique are detailed in $\S$~\ref{sec:experiment}. In $\S$~\ref{sec:modelingandsimulations} we present the formulation of our reduced-order model and describe our direct numerical simulation (DNS) framework. In $\S$~\ref{sec:results}, we compare our experiments and DNS results with our model predictions for the pertinent metrics describing drop rebound ($\S$~\ref{subsec:reboundcharacteristics}), spreading and retraction dynamics ($\S$~\ref{subsec:spreadingandretraction}), and maximum drop deformation and spreading ($\S$~\ref{subsec:DropShape}). In $\S$~\ref{sec:discussion} we synthesize our findings into new knowledge about drop rebound and contextualize it within the existing literature. Finally, we summarize our key findings in $\S$~\ref{sec:conclusions} and identify a number of open questions inspired by this work.

\section{Experiment}\label{sec:experiment}

\subsection{Experimental setup}

\begin{figure}
  \centerline{\includegraphics[width=135mm]{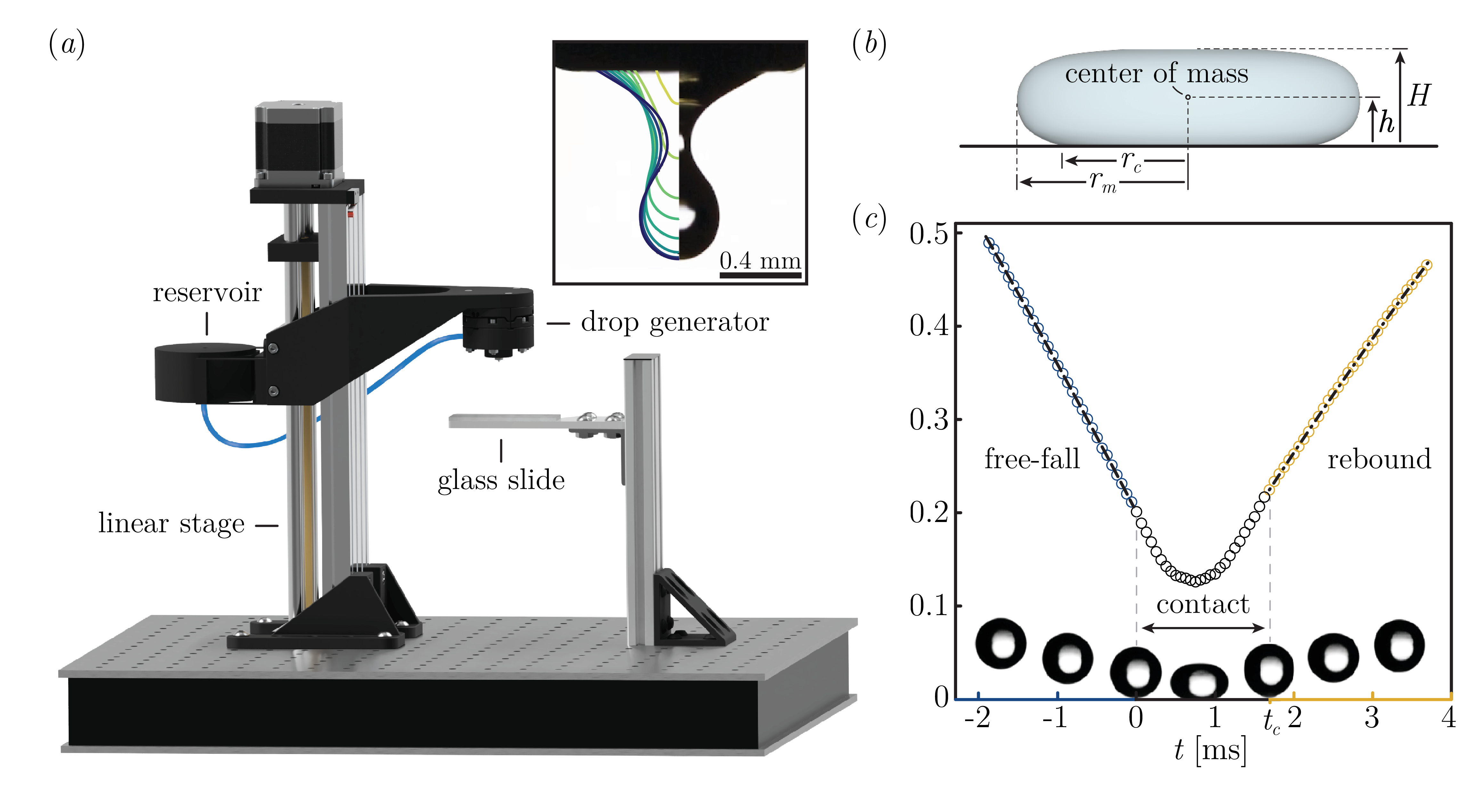}}
  \caption{($a$) A rendering of the experimental setup. The inset depicts the drop generation process with the evolution of the liquid-air interface (left half) illustrated with colored lines increasing with time from yellow (light) to blue (dark). ($b$) Schematic of an impacting drop and measured variables. ($c$) Height of the center of mass $h$ against time $t$ for a rebounding drop ($R=0.2$ mm), where $\We=0.25$, $\Oh=0.03$, and $\Bo=0.02$. Dashed and dot-dashed lines are parabolic fits to the drop position during free-fall and rebound, respectively. Gray dashed lines indicate the initial contact time $t=0$ and detachment time $t=t_c$.}
\label{fig:Fig2}
\end{figure}

We performed experiments using a 3D-printed drop generator mounted on a motorized linear stage to produce submillimeter-sized silicone oil drops with less than 1\% variability \citep{harris2015low,ionkin2018note}, as illustrated in figure~\ref{fig:Fig2}($a$). The latest design, employed for this work, is documented at \url{https://github.com/harrislab-brown/DropGen}. Among other improvements, this design allows for precise height control via a motorized linear state and is adapted to use commercially available nozzles from Fused Deposition Modeling (FDM) 3D-printers. The drop generator expels a thin liquid jet from a nozzle by deforming a piezoelectric disk and then quickly relaxes it, retracting the thin jet while surface tension drives pinch-off a single drop. By timing drop pinch-off to occur as the jet retracts, the drop is directed upward and `floats' momentarily, allowing interfacial oscillations to decay considerably. Here, we report only experiments where the ratio of the drops diameter in the horizontal and vertical directions is between 0.96 and 1.04 just before impact to minimize the effect of residual drop oscillations on the rebound metrics \citep{biance2006elasticity,yun2019spreading}. Supplemental movie 1 demonstrates drop generation. The radius of the drop $R$ and its velocity at impact $V$ were varied by using nozzles with different inner diameters and changing the distance between the nozzle and substrate, respectively. The working fluids were silicone oils (Clearco Products) with slightly varying density $\rho$ and surface tension $\sigma$, but a wide range of dynamic viscosity $\mu$, which allowed us to explore the role of drop viscosity on rebound. The ranges of the dimensional and non-dimensional parameters explored are summarized in table~\ref{tab:VariablesAndParameters}.
 
An atomically smooth, non-wetting substrate was prepared by coating a standard glass slide with a thin film of viscous silicone oil \citep{gilet2012droplets,hao2015superhydrophobic,sanjay2023drop}. To achieve this, we first coated the slide with a thick layer of viscous silicone oil $\mu=975$ g$\,$cm$^{-1}$s$^{-1}$. Then, the slide was passed beneath a fixed knife edge to thin and level the film. The mean film thickness $h_f$ was measured by weighing the slide, and was $h_f = 231 \pm 55 \ \mu$m across all experiments. This thickness corresponds to a dimensionless film thickness $\Gamma = h_f / R \approx 10^{0}$ and film mobility $\Gamma / \Oh_{f}^{1/3} \approx 10^{-2}$ that is well below the $\Gamma / \Oh_{f}^{1/3} \leq 10^{-1}$ criterion discussed earlier and thus confidently within the regime of substrate-independent drop bouncing \citep{sanjay2023drop}. After the slide was prepared it was placed in a closed environment for at least 12 hours to allow non-uniformities in the film to relax under gravity. During testing, the substrate was mounted perpendicular to the direction of gravitational acceleration to ensure the air layer between the impacting drop and substrate did not rupture prematurely due to asymmetry \citep{lo2017mechanism}.

\begin{table}
  \begin{center}
\def~{\hphantom{0}}
  \begin{tabular}{lcccc}
     Parameter \ \ \ & \ \ \ Symbol \ \ \ & \ \ \ Units (cgs) \ \ \ & \ \ \ Range \\[3pt]
      Drop radius & $R$ & cm & $0.0094-0.0967$ \\
      Impact velocity & $V$ & cm$ \, $s$^{-1}$ & $1.00-74.39$ \\
      Density & $\rho$ & g$ \, $cm$^{-3}$ & $0.850-0.975$ \\
      Surface tension & $\sigma$ & g$ \, $s$^{-2}$ & $18.7-20.8$ \\
      Dynamic viscosity & $\mu$ & g$ \, $cm$^{-1}$$ \, $s$^{-1}$ & $0.0187-0.4785$ \\
      Gravitational acceleration & $g$ & cm$ \, $s$^{-2}$ & 981 \\
      Weber number & $\We$ & $-$ & $0.0008-10.6185$ \\
      Ohnesorge number & $\Oh$ & $-$ & $0.0139-0.7865$ \\
      Bond number & $\Bo$ & $-$ & $0.0037-0.4197$ \\
  \end{tabular}
   \caption{Parameter ranges explored in our experiments.}
  \label{tab:VariablesAndParameters}
  \end{center}
\end{table}

\subsection{Experimental protocol} \label{subsec:experimentalprotocol}

To begin a set of experiments, a reservoir connected to the drop generator is filled with silicone oil, and the system is manually purged to remove any trapped air \citep{harris2015low}. Next, the pulse width controlling the piezoelectric disks actuation is tuned to produce a single drop with minimal surface oscillations at impact. These settings vary based on the working fluid and nozzle. Next, the nozzle is lowered until it rests just above the substrate, and the system height is set to zero. The nozzle is raised to the desired initial drop height, typically 1 or 2 mm, and a droplet is generated. The drop falls under the action of gravity and its motion is recorded from release until bouncing ceases. After coming to rest, an external pulse of air is applied and sweeps the drop from the substrate before it can coalesce with the film, ensuring the surface remains pristine for subsequent experiments. \edit{In the few cases where the drop merged with the viscous film, the slide was translated so the next impact occurred sufficiently far from the merge site}. The nozzle height is raised incrementally, and three to eight drops are recorded at each height. The experiments conclude once impact stresses become strong enough to pierce the intervening air layer \citep{hao2015superhydrophobic}, which occurs at $\We \approx 10$ for all $\Oh$, or drop retraction leads to droplet ejection for low $\Oh$ drops \citep{richard2000bouncing,bartolo2006singular}, as shown in supplemental movie 2. 

The silhouette of the drop was illuminated by an LED backlight and recorded using a Phantom Miro LC311 high-speed camera with a Laowa 25 mm ultra-macro lens. The frame rate ranged from 9,000 to 20,000 fps. The pixel size in the recordings was calibrated using a microscope calibration slide. The spatial resolution in our experiments varied between 0.003 and 0.008 mm depending on the size of the drop---smaller drops required higher magnification---but, in general, was approximately $0.02 R$. Recordings of each experiment were processed using MATLAB to identify the drop silhouette using Canny edge detection and calculated its center of mass by assuming the drop shape is a surface of revolution about the vertical axis passing through its centroid, and the drop has uniform density. Time was defined such that $t = 0$ when the drop ``contacts'' the substrate and $t=t_{c}$ when it detaches, where the drop and substrate are considered in ``contact'' when a portion of the drop and substrate occupy the same pixel. As noted previously, no actual contact occurs between the droplet and substrate due to the microscopic mediating air layer. During rebound, the vertical location of the center of mass $h$ and top of the drop $H$ are tracked, as well as the equatorial radius $r_m$ and contact radius $r_c$ of the drop, as defined in figure~\ref{fig:Fig2}($b$). We define $H$ as the maximum vertical distance between the substrate and any point on the surface of the drop. Consequently, when drop deformation is small, the region at the north pole of the drop is convex upward and $H$ coincides with the drop height along the vertical axis passing through is centroid. When drop deformation is large, the location along the surface of the drop where $H$ is measured shifts toward its rim.

We calculate the undisturbed drop radius $R$ by measuring the average projected area of the drop in the 30 frames preceding impact, and then solve for $R$ assuming the projected area is a circle. The center of mass $h$ during these frames is fit to a parabola and the velocity of the drop at impact is $V = dh/dt(t=0)$. Similarly, the rebound velocity is $V_r=dh/dt(t=t_{c})$, where $h$ is fit to a parabola for the 30 frames following detachment. For experiments where less than 30 frames separated detachment and subsequent impact, all intervening frames were used to calculate $V_r$. Figure~\ref{fig:Fig2}($c$) plots the center of mass of a drop $h$ against time $t$, before (blue markers), during (black markers), and after impact (yellow markers) for $\We=0.25$, $\Oh=0.03$, and $\Bo=0.02$. The vertical dashed lines indicate the initial contact time $t=0$ and detachment time $t=t_c$ of the drop, and the dashed and dot-dashed black curves are parabolic fits for $h$ during free-fall and rebound, respectively. Snapshots of the drop shape during free-fall, contact, and rebound are inset.

The elasticity of rebound can be characterized by the coefficient of restitution $\varepsilon$
\begin{equation}
\varepsilon^{2} =\frac{E_{\textrm{r}}}{E_{\textrm{i}}}= \frac{g(h(t_{c}) - R)+0.5 V_{r}^{2}}{0.5 V^2},
\label{eq:equation1}
\end{equation}
which compares the energy of the drop at rebound $E_{\textrm{r}}$ and impact $E_{\textrm{i}}$. Although $E_{\textrm{i}}$ is purely kinetic, $E_{\textrm{r}}$ has kinetic and gravitational potential contributions owing to drop deformation at detachment. In most prior studies the potential energy has been ignored and $\varepsilon = V_{r} / V$ \citep{richard2000bouncing,biance2006elasticity,de2015wettability,hao2015superhydrophobic,jha2020viscous}. Ignoring potential energy implies a spherical drop at detachment \citep{richard2000bouncing}, and is not always negligible when the kinetic energy is small for $\We<1$. A drop can be elongated at detachment when $\We<1$, as shown in the images inset in figure~\ref{fig:Fig2}($c$). Other authors defined $E_{\textrm{r}}$ as the gravitational potential energy of the drop at its maximum height after rebound, but this definition ignores possible viscous loss due to the surrounding air during free flight\citep{wang2022successive,thenarianto2023energy}. Our definition does not make any of these assumptions, providing energetically consistent and accurate measurements at low $\We$.

The uncertainty of each measurement was determined by considering the spatial and temporal resolution of our experimental setup. The temporal resolution is related to the time between successive frames $\Delta t = 0.5 \, \left(\textrm{fps}\right)^{-1}$ and the spatial resolution is calculated using the width of a pixel $L_p$, as $\Delta L = 0.5 \, \left( L_{p} \right) ^{-1}$. To estimate the uncertainty in the velocity, we first calculated the residuals between the measured center of mass height $h$ and the fitted parabola. The standard deviation of these residuals $\sigma_{\textrm{res}}$ served as a measure of the uncertainty in the position data. This uncertainty is propagated to the velocity estimate by assuming that the uncertainty in the velocity is equal to the standard deviation of the residuals, scaled by the curvature of the fitted parabola, $\delta V \approx 0.5 \, \sigma_{\textrm{res}}  \left| 2 a \right| $, where $a$ is the quadratic coefficient of the fitted parabola. Uncertainty propagation was performed using Mathematica, where first-order Taylor expansion was used for uncertainty propagation and uncertainties were assumed uncorrelated.   

\section{Modeling and simulations} \label{sec:modelingandsimulations}

\subsection{Kinematic match model formulation} \label{subsec:Formulation}

We consider a drop of incompressible fluid of constant density $\rho$, dynamic viscosity $\mu$, and surface tension coefficient $\sigma$ impacting normally on a perfectly hydrophobic (non-wetting), horizontal solid substrate. Before impact, the droplet is spherical with radius $R$ and the surface of the droplet is stationary relative to its center of mass. During impact the drop is ``pressed" to the substrate rather than in true physical ``contact" with it; however, we will use the later in our discussion for consistency and connection with the prior literature. The criteria for contact and its implication on the model results are discussed in Appendix~\ref{appA}.

We introduce spherical coordinates $(\zeta',\theta,\phi)$ with $\theta\in\left[0,\pi\right]$ and $\phi\in\left[0,2\pi\right]$ (where~$'$ indicates a dimensional variable) to describe the free surface of the drop. Here, $\theta$ is measured relative to the direction of gravity, and the origin is at the center of mass of the drop. Due to the assumed axisymmetry, the azimuthal angle $\phi$ can be ignored, consistent with experiments for all cases considered here. Additionally, we define the associated Cartesian system of coordinates $(x',z')$, as shown in figure~\ref{fig:model_graphic}($a$) for an impacting drop. This formulation allows the smooth drop shape to be discretized, as shown in figure~\ref{fig:model_graphic}($b$), and permits numerical approximations that enable efficient simulation of drop rebound across a wide range of $\We$.

\subsubsection{Drop motion} \label{subsubsec:dropmotion}

 The drop is subject to gravitational acceleration $g\mathbf{ \hat{z}}$, and the substrate is at a distance $h'$ from its center of mass. At time $t'=0$, $h'=R$, and the center of mass of the drop moves towards the substrate with velocity $V$. The fluid motion within the drop is governed by the Navier-Stokes equations for incompressible flow, with standard kinematic and dynamic boundary conditions. The normal impact on the substrate is modeled by an evolving pressure distribution over the contact area $S \left( t \right)$, where the drop surface and substrate coincide in the model.
 
\begin{figure}
  \centering
  \includegraphics[width=135mm]{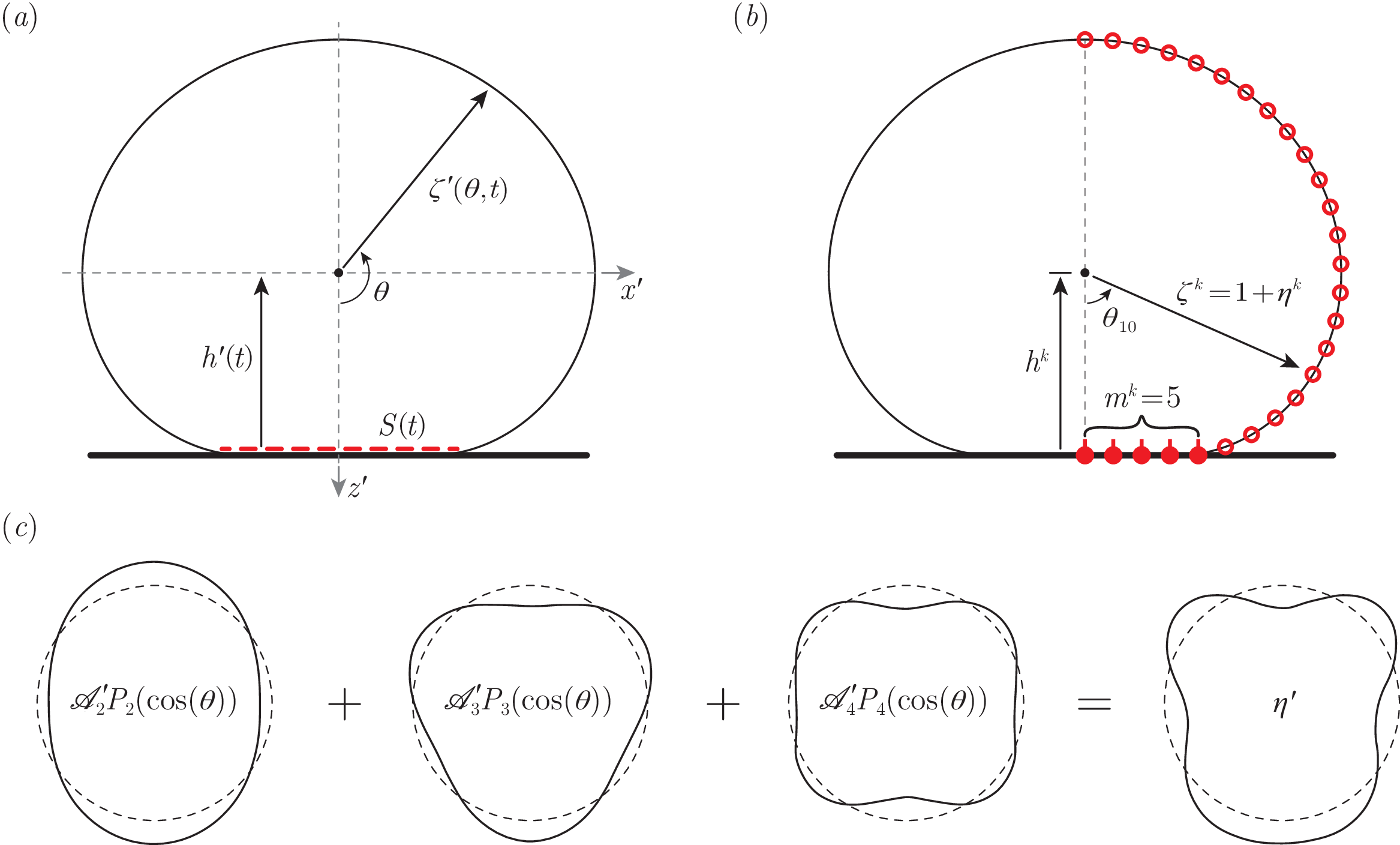}
  \caption{($a$) Schematic of an axisymmetric drop impacting a rigid substrate, where $\zeta ' \left( \theta , t \right)$ is the dimensional interface, $h'\left( t \right)$ the height of the center of mass of the drop, and $S \left( t \right)$ the contact surface. ($b$) Angular discretization of the right hemisphere of an axisymmetric drop during impact. Filled circles represent mesh points where the drop and substrate touch, while hollow circles denote points along the liquid-gas interface. ($c$) An example of shape reconstruction through Legendre synthesis for modes $l=2-4$.}
\label{fig:model_graphic}
\end{figure}

 Following \cite{alventosa2023inertio}, we linearize the governing equations and represent the surface deformation of the drop $\eta '$ as a sum of axisymmetric spherical harmonic modes:
\begin{equation}
    \eta' \coloneqq \zeta'-R = \sum\limits_{l=2}^\infty {\mathscr{A'}_lP_l(\cos(\theta))}.
\end{equation}
Here, $P_l$ are Legendre polynomials, and $\mathscr{A'}_l$ denotes the time-dependent amplitude of the $l^{\text{th}}$ mode. The $l=0$ mode is excluded due to incompressibility.  The $l=1$ mode is also excluded in the present formulation due to the choice of the drop's center of mass as the origin, since the $l=1$ mode shifts the center of mass. Figure~\ref{fig:model_graphic}($c$) demonstrates the synthesis of Legendre polynomials for modes $l=2-4$ into a sample deformed drop shape. Similarly, the contact pressure $p'$ exerted by the substrate on the drop is expressed as
\begin{equation}
    p' = \sum\limits_{l=0}^\infty {\mathscr{B'}_lP_l(\cos(\theta))},
\end{equation}
where $\mathscr{B'}_l$ is the time-dependent amplitude of the $l^{\text{th}}$ Legendre mode.
 
The evolution of the drop interface is obtained by applying linearized kinematic conditions on the surface in the weakly damped and small-amplitude regime \citep{Lamb1924,miller1968oscillations,Phillips_Milewski_2024}
\begin{equation}\label{eqn:dimensional_dot_A_l}
    \dot{\mathscr{A}}_l' = \mathscr{U}_l', \qquad \text{ for } l = 2,3,\ldots, \infty;
\end{equation}
\begin{equation}\label{eqn:dimensional_dot_U_l_v01}
    \dot{\mathscr{U}}_l'
    =
    -\frac{\sigma l \left(l+2\right)\left(l-1\right)}{\rho R^3}\mathscr{A}_l'
    -2(2l+1)(l-1)\frac{\mu}{\rho R^2}\mathscr{U}_l'
    -\frac{l}{\rho R}\mathscr{B}_l', \qquad \text{ for } l = 2,\ldots,\infty,
\end{equation}
where $\mathscr{U}_l'$ is the amplitude of the $l^\text{th}$ Legendre mode of the surface's velocity, $\dot{\zeta'}$. Equations \eqref{eqn:dimensional_dot_A_l} and \eqref{eqn:dimensional_dot_U_l_v01} are subject to 
\begin{equation}
    \mathscr{A}_l'(t=0) = 0, \qquad \text{ for } l = 2,\ldots,\infty;
\end{equation}
\begin{equation}
    \mathscr{U}_l'(t=0) = 0, \qquad \text{ for } l = 2,\ldots,\infty.
\end{equation}

The height of the center of mass is naturally governed by Newton's second law of motion,
\begin{equation}
    \dot{h}' = v',
\end{equation}
\begin{equation}
    \dot{v}' = -g + \frac{1}{m}\frac{4\pi R^2}{3}\mathscr{B}_1',
\end{equation}
subject to
\begin{equation}
    h'(t=0) = R,
\end{equation}
\begin{equation}
    v'(t=0) = -V,
\end{equation} where $m=\frac{4\pi}{3}\rho R^3$ is the mass of the drop and $g$ is gravitational acceleration.

\subsubsection{The kinematic match} \label{subsubsec:thekinematicmatch}

The motion of the drop's surface must satisfy additional compatibility conditions. Namely, when the substrate exerts pressure on the drop, a circular contact area $\theta\leq\theta_c$ forms, which must lie on the $z'=0$ plane. That is
\begin{equation}
    \sum\limits_{l=2}^\infty {\mathscr{A}_l'P_l(\cos(\theta))} = \frac{h'}{\cos(\theta)}-R, \qquad \text{ for } \theta \leq \theta_c.
\end{equation}
Outside the contact area, the drop surface must be strictly above the solid surface,
\begin{equation}
    \sum\limits_{l=2}^\infty {\mathscr{A}_l'P_l(\cos(\theta))} > \frac{h'}{\cos(\theta)}-R, \qquad \text{ for } \theta > \theta_c,
\end{equation}
and the contact pressure there must be null, that is,
\begin{equation}
    \sum\limits_{l=0}^\infty {\mathscr{B}_l'P_l(\cos(\theta))} = 0, \qquad  \text{ for } \theta > \theta_c.
\end{equation}
Finally, at the boundary of the contact area, the drop surface must be tangent to the substrate, 
\begin{equation}
\frac{\partial}{\partial \theta}\left[\left(R+\sum\limits_{l=2}^\infty {\mathscr{A}_l'P_l(\cos(\theta))}\right)\cos(\theta)\right]_{\theta=\theta_c} = 0,
\end{equation}
which captures the non-wetting nature of the surface. Our formulation results in a fully self-contained model without fitting parameters, which we can now non-dimensionlize.

\subsubsection{Non-dimensionalization} \label{subsubsec:nondimensionalization}

The characteristic length, mass, and time are taken to be the undeformed droplet radius $R$, mass scale $\rho R^{3}$, and the inertio-capillary time $t_{\sigma}= \left(\left( \rho R^{3} \right) / \sigma \right)^{1/2}$, respectively. We also define the Bond number $\Bo$, Weber number $\We$, and Ohnesorge number $\Oh$ as in equation (\ref{eqn:dimensionlessparameters}). This process leads to the dimensionless equations:
\begin{equation}\label{eqn:nd_dot_A_l_v01}
    \dot{\mathscr{A}}_l = \mathscr{U}_l, \qquad \text{ for } l = 2,\ldots, \infty;
\end{equation}
\begin{equation}\label{eqn:nd_dot_U_l_v01}
    \dot{\mathscr{U}}_l 
    =
    -l \left(l+2\right)\left(l-1\right)\mathscr{A}_l
    -2(2l+1)(l-1) \Oh \, \mathscr{U}_l
    -l \mathscr{B}_l, \qquad \text{ for } l = 2,\ldots, \infty;
\end{equation}
\begin{equation}\label{eqn:nd_dot_h_v}
    \dot{h} = v;
\end{equation}
\begin{equation} \label{eqn:nd_dot_v}
    \dot{v} = - \Bo + \mathscr{B}_1;
\end{equation}
which are subject to 
\begin{equation}
    \mathscr{A}_l(t=0) = 0, \qquad \text{ for } l = 2,\ldots, \infty;
\end{equation}
\begin{equation}
    \mathscr{U}_l(t=0) = 0, \qquad \text{ for } l = 2,\ldots, \infty;
\end{equation}
\begin{equation}
    h(t=0) = 1;
\end{equation}
\begin{equation}
    v(t=0) =- \sqrt{\We};
\end{equation}
\begin{equation}\label{eqn:km_contact_amplitudes}
    \sum\limits_{l=2}^\infty {\mathscr{A}_lP_l(\cos(\theta))} = \frac{h}{\cos(\theta)}-1, \qquad \text{ for } \theta \leq \theta_c;
\end{equation}
\begin{equation}\label{eqn:km_non-contact_amplitudes}
    \sum\limits_{l=2}^\infty {\mathscr{A}_lP_l(\cos(\theta))} > \frac{h}{\cos(\theta)}-1, \qquad \text{ for } \theta > \theta_c;
\end{equation}
\begin{equation}\label{eqn:km_pressure_amplitudes}
    \sum\limits_{l=0}^\infty {\mathscr{B}_lP_l(\cos(\theta))} = 0, \qquad \hspace{14 mm} \text{ for } \theta > \theta_c;
\end{equation}
\begin{equation}\label{eqn:km_contact_derivative}
    \frac{\partial}{\partial \theta}\left[\left(1+\sum\limits_{l=2}^\infty {\mathscr{A}_lP_l(\cos(\theta))}\right)\cos(\theta)\right]_{\theta=\theta_c} = 0.
\end{equation}

We solve this system of equations by first applying numerical approximations, detailed in appendix~\ref{appB}. Briefly, these include truncating the expansions of the surface shape, velocity, and pressure distribution with a spectral mesh consisting of Legendre polynomials up to degree $L=90$ and approximating time derivatives using a first-order implicit Euler method. All simulations used to generate the results in the remaining figures of this paper adopt this value of $L$. A schematic of the angular mesh is shown in figure~\ref{fig:model_graphic}($b$). The result of these approximations is a discrete optimization problem, which we solve using an adaptive timestep and error minimization scheme, as described in appendix~\ref{appC}. This approach accurately solves the system at reduced computation cost relative to DNS, with the longest simulations completing in $\mathcal{O}(10)$ CPU minutes and a typical simulation time of less than 1 CPU minute.

\subsection{Direct numerical simulation} \label{subsec:dns}

We develop a computational framework based on the Basilisk open-source code structure \citep{popinet2009accurate,popinet2015quadtree} as a comparative benchmark for the mathematical model described in Section~\ref{subsec:Formulation}, and to interrogate quantities difficult to visualize experimentally. The fully nonlinear implementation is used to resolve the time-dependent Navier-Stokes equations in both liquid and gas phases, with physical properties of the latter taken to be the values of air at room temperature i.e. $\rho_g = 0.00121$ g$ \, $cm$^{-3}$ and $\mu_g = 0.000181$ g$ \, $cm$^{-1}$$ \, $s$^{-1}$, in an attempt to replicate the experimental conditions. \edit{Basilisk employs the volume-of-fluid method (\cite{hirt1981volume, scardovelli1999direct}) for the implicit representation of the interface describing the transition between the different fluid regions, with a so-called color function $f$ taking the value $1$ inside the liquid and $0$ inside the gas regions, respectively, and values $0 < f < 1$ representing computational cells cut by a fluid-fluid interface.} The adaptive mesh refinement and parallelization capabilities of the implementation make it a particularly good candidate for this multi-scale impact problem. The code base for bouncing problems has been constructed and systematically validated for several years in our extended group, covering fluid-structure interaction \citep{galeano2021capillary} and liquid-liquid impact \citep{alventosa2023inertio}, with comprehensive numerical campaigns in the wider community \citep{sanjay2023does,sanjay2023drop} providing key insights into related regimes.

\begin{figure}
  \centering
  \includegraphics[width=135mm]{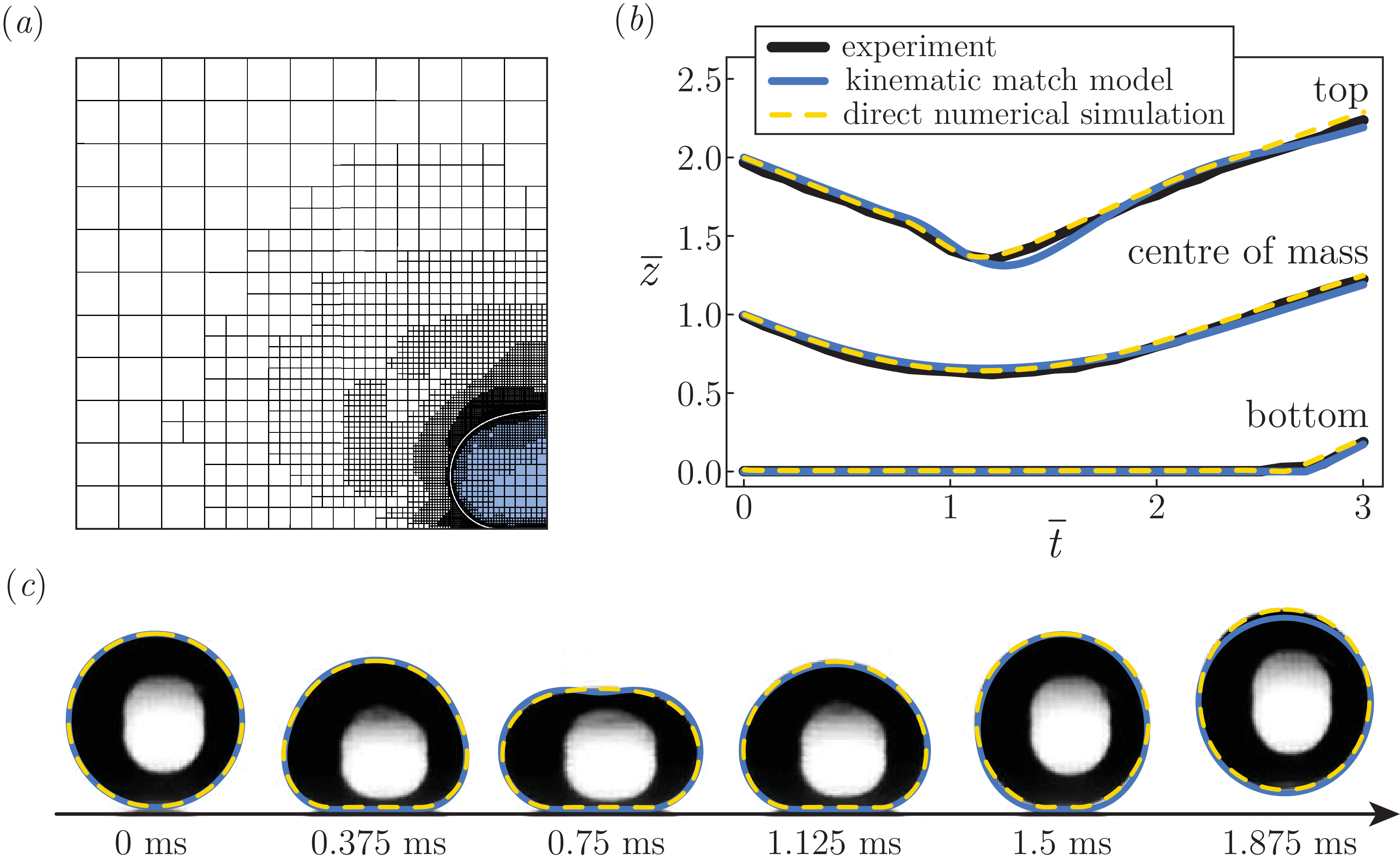}
  \caption{($a$) Visualization of the computational domain comprising of liquid (blue) and gas regions (white), and underlying adaptive grid cell structure in the direct numerical simulation, for a typical $\We=0.253$ impact taken at the point of maximal radius deformation $t=0.746$ ms. ($b$) Dimensionless distance above the substrate $\bar{z} = z / R$ against time $\bar{t} = t / t_{\sigma}$ for the top, bottom, and center of mass of a drop for $\We= 0.253$, $\Bo = 0.02$ and $\Oh = 0.03$. Black solid lines represents experiment, while blue solid lines and yellow dashed lines represent the KM model and direct numerical simulations, respectively. ($c$) Overlay of the predicted shape from the KM model (solid blue) and direct numerical simulations (dashed yellow) with experimental images corresponding to ($b$). The drop radius is $R=0.203$ mm and the time interval between images is 0.375 ms.}
\label{fig:overlay}
\end{figure}

The axisymmetric computational domain set up in cylindrical coordinates measures $8R$ across both ($r-$ and $z-$) spatial dimensions. To ensure sufficient resources are dedicated to the delicate bouncing dynamics, we use a strong grid cell refinement level with a minimal grid cell size of approximately $1/20\ \mu$m, verified to be robust across our parameter regime of interest. This stringent resolution specification was particularly relevant at the lowest Weber numbers studied herein, where the implementation requires most resilience in view of increased numerical stiffness. Adaptivity was tailored around the evolving location of the drop interface and changes in magnitude of the velocity components in the flow, as illustrated in figure~\ref{fig:overlay}($a$). Special care was taken during runtime via stringent projection solver tolerances and increased number of internal iterations set for the multigrid cycle to ensure a high degree of accuracy. Furthermore, a zero Dirichlet boundary condition for the color function \edit{$f$} was prescribed at the level of the solid surface to prevent touchdown and contact, with all tests resulting in the drop rebounding or resting on the surface. We employ a criterion based on the drop interface crossing a threshold level of $0.02R$ above the substrate to define the initialization and end of contact as discussed in Appendix~\ref{appA}. This criterion enables us to build a consistent comparative framework with the model and experiment. Under these specifications, a typical run is characterised by $\mathcal{O}(10^5)$ grid cells and $\mathcal{O}(10-100)$ CPU hours of runtime, executed across multiple cores on local high-performance computing architectures. Mesh-independent results in our target metrics have been observed under the conditions above and reported in the sections to follow. All associated code, including setup scripts and post-processing functionality, are provided in the form of a GitHub repository at \url{https://github.com/rcsc-group/BouncingDropletsSolidSurface}.

\subsection{Validation} \label{subsec:validation}

Beyond verification and resolution studies across our parameter space of interest, we also validate our theoretical approaches with experimental data. A typical case illustrating a comparison between the kinematic match (KM) model and DNS is the rebound of a drop with radius $R=0.203$ mm for $\We=0.253$, $\Bo=0.02$, and $\Oh=0.03$. Figure~\ref{fig:overlay}($b$) plots the normalized distance from the substrate $\bar{z}=z/R$ of the top, bottom, and center of mass of the drop against time $\bar{t}=t/t_{\sigma}$. The experimental trajectory is shown in black and aligns well with predictions from the KM model (blue solid line) and DNS (yellow dashed line) for all three locations. Figure~\ref{fig:overlay}($c$) shows overlays of the predicted drop shape from the KM model (solid blue line) and DNS (yellow dashed line) onto experimental images, demonstrating strong visual agreement. Supplemental movie 3 shows this rebound event with overlays. Additionally, we note that the accuracy of the KM model improves on a relative basis as $\We$ decreases, as demonstrated in supplemental movie 4 for a similar drop at $\We=0.023$. This low-$\We$ regime is of high consistency with the underlying simplifying assumptions, one in which resources can also be deployed very efficiently given smaller interface deformations. By contrast, we find that the DNS framework struggles for sufficiently low values of $\We$ (which we identified to be of $\mathcal{O}(10^{-2})$ or lower) without significant customization or computational re-design, as both numerical stiffness and parasitic capillary currents become progressively more difficult to contain accurately. This makes the cross-validation and interplay with the KM model even more constructive as our investigation reaches distinguished limits of our target parameter regime.

\section{Results}\label{sec:results} \label{sec:results}

We performed experiments over a wide range of $\We$, $\Oh$ and $\Bo$, as highlighted in supplemental movies 5, 6, and 7. In this section, we compare our results with direct numerical simulation data and the KM model. We first investigate the metrics defining rebound, the coefficient of restitution $\varepsilon$ and contact time $t_c$. Next, we analyze the evolution of the drop shape and timescales associated with spreading and retraction. Last, we show how the maximum deformation of the drop changes over our parameter space and compare with the KM model and energetic arguments.

\subsection{Rebound metrics} \label{subsec:reboundcharacteristics}

\begin{figure}
  \centering
  \includegraphics[width=135mm]{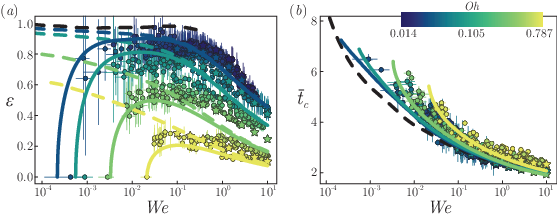}
  \caption{($a$) Coefficient of restitution $\varepsilon$ against Weber number $\We$ for drops with varying $\Oh$. Circles denote experiments and stars represent DNS results for $\Bo=0.02$, a typical value in experiment. Solid and dashed lines are the KM model for $\Bo=0.02$ and $\Bo=0$, where the black dashed line shows the inertio-capillary limit ($\Oh=0$; $\Bo = 0$). ($b$) Dimensionless contact time $\bar{t}_c = t_{c} / t_{\sigma}$ against $\We$ for experiments, DNS, and model predictions from ($a$). The color bar maps $\Oh$ to color on a logarithmic scale.}
\label{fig:rebound_characteristics}
\end{figure}

The coefficient of restitution $\varepsilon$ quantifies the energy recovered during rebound, and behaves differently at low and high $\We$. Figure~\ref{fig:rebound_characteristics}($a$) shows $\varepsilon$ against $\We$, for various $\Oh$ spanning two orders of magnitude (as denoted by color). Here, circle markers represent experiments and star markers represent DNS results for $\Bo=0.02$, the most common value in our low-$\We$ experiments. The solid and dashed lines are predictions from the KM model for $\Bo=0.02$ and $\Bo=0$, respectively. The intercept of a curve with the $x$-axis signifies the cessation of rebound. The KM model predictions for the experimental trends at low $\We$ agree reasonably well with experiment and produce robust predictions even at very low $\We$ where the DNS struggles. At high $\We$, the DNS predictions agree better with experiment, where the assumptions of the kinematic model are less justified. The complementary modeling approaches allows us to confidently span the full region accessed by experiment, and beyond.

For $\We \gtrapprox 0.5$, $\varepsilon$ decreases as $\We$ increases for each $\Oh$ in agreement with prior literature \citep{biance2006elasticity,jha2020viscous}. In addition, the black dashed line shows the prediction of the KM model in the strict inertio-capillary limit ($\Oh = 0$; $\Bo = 0$). Even in the complete absence of gravity and viscosity, our model predicts a similar roll-off at high $\We$ due to energy transfer to droplet oscillations not accounted for in the definition of $\varepsilon$.  Curiously, this high-$\We$ roll-off is notably absent for the analogous case of droplet-bath rebounds, where for low $\Oh$ and $\Bo$, the coefficient of restitution levels off at approximately $0.3$ \citep{zhao2011transition,wu2020small,alventosa2023inertio}.

Below $\We \approx 0.5$, $\varepsilon$ becomes nearly $\We$-independent across roughly two orders of magnitude when $\Oh$ is small. As $\Oh$ increases, the influence of $\We$ on $\varepsilon$ increases. As $\We$ decreases, $\Bo$ effects cause $\varepsilon \rightarrow 0$ at a critical Weber number $\We_c$ that increases with $\Oh$. For the experimental data presented, $\varepsilon = 0$ represents the highest $\We$ experiment where rebound was not detected (for each viscosity). In related work for low $\We$ rebound on hydrophobic surfaces, a similar roll-off leading to bounce cessation was attributed to contact line friction \citep{wang2022successive,thenarianto2023energy}, wholly absent in the current experiments and models. To investigate the physical origin of this threshold in our fully non-wetting scenario, we performed additional KM simulations in the absence of gravity (i.e. $\Bo=0$, shown as the colored dashed curves).  For the parameters considered here, $\We_c$ is eliminated in the absence of gravity, effectively extending bouncing to lower $\We$. Gravity shifts the static equilibrium height of the drop's center of mass below one radius above the surface, representing a higher barrier for complete rebound when energy is lost to viscosity or transferred to droplet oscillations during contact.  In the low $\We$ regime, the KM model predicts nearly perfect elastic collisions ($\varepsilon\approx 1$) in the inertio-capillary limit ($\Oh=0$; $\Bo=0$).

The contact time $t_c$ follows the simple scaling $t_{c} \sim t_{\sigma}$ for moderate and high $\We$ \citep{richard2002contact} but depends on $\We$ and $\Oh$ at low $\We$. Figure~\ref{fig:rebound_characteristics}($b$) shows the nondimensional contact time $\bar{t}_{c} = t_{c} / t_{\sigma}$ against $\We$ for drops of varying $\Oh$. The results correspond with those in figure~\ref{fig:rebound_characteristics}($a$); thus, plot markers, line types, and colors hold the same meaning. For all $\Oh$, the contact time approaches a constant as $\We$ increases, in agreement with prior literature \citep{okumura2003water}. As $\We$ decreases, $t_c$ increases at an $\Oh$-dependent rate; the rate of increase is greater for larger $\Oh$. Thus, the effects of viscosity on contact time are most influential at low $\We$ and may alter the utility of drops in a number of applications involving heat transfer \citep{shiri2017heat}, for instance. The KM model agrees well here for all $\We$ and $\Oh$, and the inertio-capillary limit sets a lower bound on the contact time and an upper bound on the coefficient of restitution for each $\We$, similar to the case for droplet-bath rebound \citep{alventosa2023inertio}.

\subsection{Spreading and retraction} \label{subsec:spreadingandretraction}

\begin{figure}
  \centering
  \includegraphics[width=135mm]{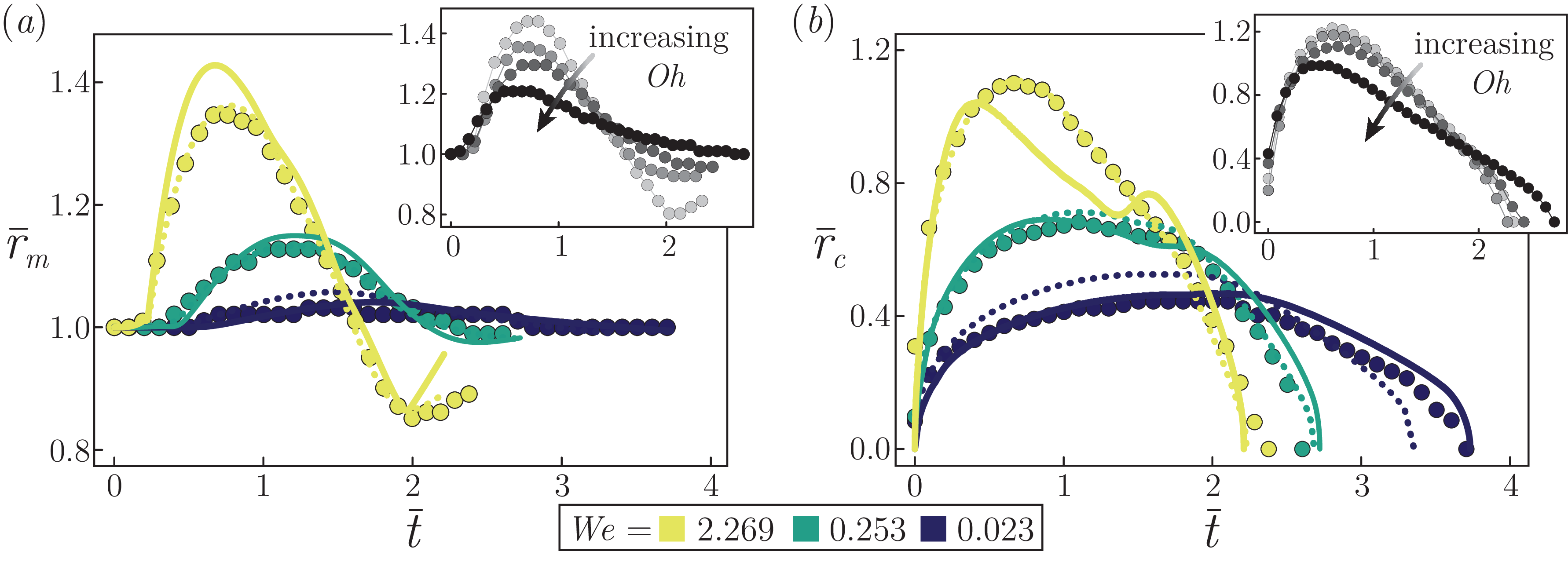}
  \caption{($a$) Dimensionless equatorial radius $\bar{r}_{m} = r_{m} / R$ against dimensionless time $\bar{t}=t / t_{\sigma}$ for experiments (markers), DNS (dotted lines), and the KM model (solid lines) for $\We = 2.269$ (light yellow), $\We = 0.253$ (green), and $\We = 0.023$ (dark blue), where $\Bo=0.02$ and $\Oh=0.03$ for all curves. ($b$) Dimensionless contact radius $\bar{r}_{c} = r_{c} / R$ against dimensionless time $\bar{t}=t / t_{\sigma}$, where line, marker type, and color are the same as ($a$). The insets show experimental results for drops of similar $\We$ ($2.5-3.5$) and $\Bo$ ($0.16-0.24$) for $\Oh=$ 0.03, 0.08, 0.29, and 0.7, where marker darkness indicates higher $\Oh$.}
\label{fig:spreading_and_retraction}
\end{figure}

The shift in rebound behavior between low and high $\We$ revealed in $\S$~\ref{subsec:reboundcharacteristics} coincides with a shift in the spreading and retraction dynamics. Figure~\ref{fig:spreading_and_retraction}($a$) shows the non-dimensional equatorial radius $\bar{r}_{m} = r_{m} / R$ against non-dimensional time $\bar{t} = t / t_{\sigma}$ during contact for $\We=0.023$ (dark blue), $\We=0.253$ (green), and $\We=2.269$ (light yellow), with $\Bo=0.02$ and $\Oh = 0.03$. Circles represent experimental measurements, while dotted and solid lines denote DNS and the KM model results, respectively. In all cases, inertia drives spreading, causing $\bar{r}_{m}$ to increase after an initial period of localized deformation where $\bar{r}_{m} \approx 1$. As $\We$ increases, the drop expands quicker and $\bar{r}_{m}$ reaches a higher maximum. After the drop reaches its maximum extent, surface tension drives retraction, decreasing $\bar{r}_{m}$. For $\We = 0.023$, retraction mirrors spreading, and $\bar{r}_{m}$ evolves symmetrically, consistent with quasi-static-like behavior \citep{molavcek2012quasi,chevy2012liquid}. When $\We = 2.269$, the two phases are dissimilar, and $\bar{r}_{m}$ overshoots $\bar{r}_{m} = 1$ during retraction, leading to apparent underdamped oscillations. The inset plot shows the evolution of $\bar{r}_{m}$ for drops of similar $\We$ and $\Bo$ but varying $\Oh$, where darker marker shade indicates higher $\Oh$. Increasing viscous dissipation can eliminate the appearance of underdamped behavior; however, it is not sufficient to recover the symmetry between spreading and retraction. Instead, higher $\Oh$ shortens the spreading phase, reduces the maximum $\bar{r}_{m}$, and lengthens the retraction phase, thereby increasing asymmetry in the spreading and retraction process.

The increased asymmetry between spreading and retraction as $\We$ increases is also observed in the nondimensional contact radius $\bar{r}_{c} = r_{c} / R$. Figure~\ref{fig:spreading_and_retraction}($b$) shows $\bar{r}_{c} = r_{c} / R$ against nondimensional time $\bar{t} = t / t_{\sigma}$ for $\We=0.023$ (dark blue), $\We=0.253$ (green), and $\We=2.269$ (light yellow). The data and model predictions correspond with those in figure~\ref{fig:spreading_and_retraction}($a$); thus, plot markers, line types, and colors hold the same meaning. Unlike $\bar{r}_{m}$, there is no delay in the response of $\bar{r}_{c}$ to impact as $\bar{t} = 0$ is defined as the moment the drop and substrate ``contact''. As $\We$ increases, the maximum $\bar{r}_c$ increases, while the time to reach it decreases. For $\We = 0.023$, the spreading and retraction of $\bar{r}_c$ are symmetric, whereas at $\We = 2.269$, they are asymmetric, qualitatively resembling the behavior of $\bar{r}_m$ during rebound. The inset shows the evolution of $\bar{r}_c$ for drops of similar $\Bo$ and $\We$, but increasing $\Oh$, as indicated by darker shaded markers. Similar to $\bar{r}_m$, increasing $\Oh$ shortens the spreading phase by reducing the maximum $\bar{r}_c$, and extending the retraction phase. The asymmetry between spreading and retraction motivates a detailed look at their relationship across all experiments.

\begin{figure}
  \centering
  \includegraphics[width=95mm]{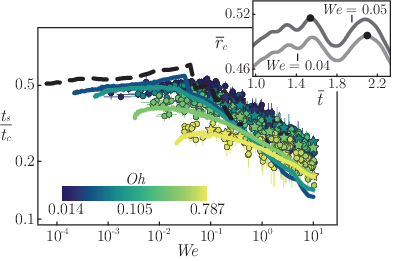}
  \caption{The relative spreading time $t_{s} / t_{c}$ against Weber number $\We$ for drops with varying $\Oh$. Circles represent experiment and stars denote DNS for $\Bo=0.02$. Solid lines are predictions from the KM model for $\Bo = 0.02$, and the black dashed line shows the inertio-capillary limit ($\Oh = 0$; $\Bo = 0$). The inset shows the predicted dimensionless contact radius $\bar{r}_c$ against dimensionless time $\bar{t} = t / t_{\sigma}$ for $\We=0.04$ and $\We=0.05$, where $\Bo = 0$ and $\Oh = 0$. The black dots show when $\bar{r}_c$ is maximum.}
\label{fig:trcPlot}
\end{figure}

The spreading time $t_s$ is the time from when the drop first contacts the substrate until $r_c$ reaches its maximum value $R_c$, and is proportional to the retraction time $t_r$ at low $\We$. In figure~\ref{fig:trcPlot} we show the relative spreading time $t_{s} / t_{c}$, where the contact time is the sum of the spreading and retraction time $t_{c} = t_{s} + t_{r}$, against $\We$ for all $\Oh$. Solid lines represent predictions from the KM model for $\Bo=0.02$, with the black dashed curve representing the inertio-capillary limit ($\Bo=0$; $\Oh=0$). When $\We \gtrapprox 0.5$, drops spread faster than they retract, and increasing $\We$ increases the difference between $t_s$ and $t_r$. However, the time the drop spends spreading relative to retracting is nearly $\We$-independent for $\We \lessapprox 0.5$, until the drop ceases to bounce. This trend holds for all viscosities, with $\Oh$ limiting the maximum of $t_{s} / t_{c}$. Although this maximum approaches 0.5 as $\We \rightarrow 0$ in the inertio-capillary limit, it surpasses it and peaks near $\We = 0.04$. The inset in figure~\ref{fig:rebound_characteristics} shows the evolution of the dimensionless contact radius $\bar{r}_c$ near its maximum, which is denoted by a black dot. Oscillations due to capillary waves cause an apparent jumpiness in $t_s$ at low $\Oh$. In particular, as $\We$ decreases from $\We = 0.05$ to $\We= 0.04$, the maximum contact radius suddenly switches between its two largest peaks, producing the apparent sudden rise in $t_s$. These results reveal that the symmetry of the spreading and retraction process coincides with nearly-elastic rebound metrics that are independent of $\We$ \edit{in the inertio-capillary limit}. In contrast, the emergence of asymmetry coincides with the onset of $\We$-dependent trends for $\varepsilon$ while increasing $\Oh$ similarly decreases $t_{s}/t_{c}$ and $\varepsilon$ across all $\We$. As in the prior rebound measurements, the KM model performs best for low $\We$ where as the DNS excels at higher $\We$ for the spreading and retraction metrics presented throughout this section.

\subsection{Maximum spreading and deformation} \label{subsec:DropShape}

We next examine how drop deformation depends on the governing dimensionless parameters and relate deformation and spreading. To do this, we define the maximum equatorial deformation $\beta - 1$, where $\beta = R_{m} / R$ is the \edit{nondimensional} maximum equatorial radius \edit{and $R_{m}=\textrm{max}(r_{m}(t))$}. Similarly, we define maximum vertical deformation $1 - \alpha$, where $\alpha = H_{min} / 2R$ and \edit{$H_{min} = \textrm{min}(H(t))$}. Lastly, we define the maximum spreading radius as $\beta_{c} = R_{c} / R$, \edit{where $R_{c} = \textrm{max}(r_{c}(t))$.}

\subsubsection{Energy argument} \label{subsubsec:energyarguments}
In addition to the kinematic match model, we can compare the drop spreading and deformation to a simpler energetic argument by assuming the drop principally deforms in the $l=2$ mode in the low-$\We$ regime.  Under this assumption, the instantaneous drop shape is fully defined by the coefficient $\mathscr{A}_2^{\prime}$, with $r_{m}=R-\frac{1}{2}\mathscr{A}_2^{\prime}$ and $h=2R+2\mathscr{A}_2^{\prime}$. 

The anomalous surface energy associated with drop deformations of the $l$-th mode (for $l\geq2$) can be expressed as
\begin{equation}
E_l=2 \pi \sigma \frac{(l+2)(l-1)}{2 l+1} \mathscr{A}_l^{\prime 2},
\end{equation}
with the surface energy in the $l=2$ mode being $E_2=\frac{8}{5} \pi \sigma \mathscr{A}_{2}^{\prime 2}$ \citep{molavcek2012quasi}. Neglecting gravity ($\Bo^{2} \ll \We$) and viscous dissipation ($\Oh\ll1$), \edit{and assuming the initial kinetic energy of the droplet ($E_V=\frac{2}{3}\pi\rho R^3 V^2$) is fully converted into surface energy of the $l=2$ mode (i.e. $E_V=E_2$), one finds}
\begin{equation}
\beta-1=\sqrt{\frac{5}{48}} \We^{1 / 2}  \label{Beta_We}
\end{equation}
 and 
 \begin{equation}
1-\alpha=\sqrt{\frac{5}{12}} \We^{1 / 2}, \label{Alpha_We}
 \end{equation}
\edit{using the relations $\mathscr{A}_{2}^{\prime}=2R\left(1-\beta\right)$ and $\mathscr{A}_{2}^{\prime}=R\left(\alpha-1\right)$, respectively, and assuming $\mathscr{A}_{2}^{\prime}\leq0$ (i.e. $R_m\geq R$, $H_{min}\leq 2R$) in the deformed state.} The result for $\beta - 1$ \edit{(equation \ref{Beta_We}) is} identical to that derived by \citet{richard2000bouncing} using similar arguments.   This derivation also suggests a geometric relationship between $\beta$ and $\alpha$: 
 \begin{equation}
1-\alpha=2(\beta-1). \label{Alpha_Beta}
 \end{equation}
 To connect this deformation to the contact radius $r_c$, we appeal to a quasi-static argument expected to be valid for $\We\ll 1$ \citep{chevy2012liquid,molavcek2012quasi}.  The force associated with the creation of additional surface area in the $l=2$ mode can be expressed as the derivative of the surface energy, $F_2=-\frac{dE_2}{d\mathscr{A}_2^{\prime}}=-\frac{16}{5}\pi\sigma\mathscr{A}_2^{\prime}$.  Following \citet{chevy2012liquid}, we assume small deformations such that the capillary pressure inside the drop is approximately $2\sigma/R$, and thus the net vertical force associated with the `contact' region is $F_{\sigma}=\frac{2\sigma}{R}\pi r_c^2$.  Balancing these forces, one arrives at the relationship $\frac{r_c^2}{R^2}=-\frac{8}{5} \frac{\mathscr{A}_2^{\prime}}{R}$.  This result can be combined with the prior relations to arrive at the prediction
\begin{equation}
\beta_c^2=\frac{16}{5}(\beta-1) \label{BetaC_Beta}
\end{equation}
or 
\begin{equation}
\beta_c=\left(\frac{16}{15}\right)^{1/4} \We^{1 / 4}. \label{BetaC_We}
\end{equation}

\edit{Note that equations~(\ref{Beta_We})–(\ref{BetaC_We}) have limited predictive power compared with the KM model, which captures viscous effects and the complete evolution of the drop shape. Nevertheless, the favorable comparison between the simple energy argument and KM model elucidates the dominant mode of deformation and motivates the parametric scalings.}

\subsubsection{Comparisons} \label{subsubsec:comparisions}

\begin{figure} 
  \centering
  \includegraphics[width=135mm]{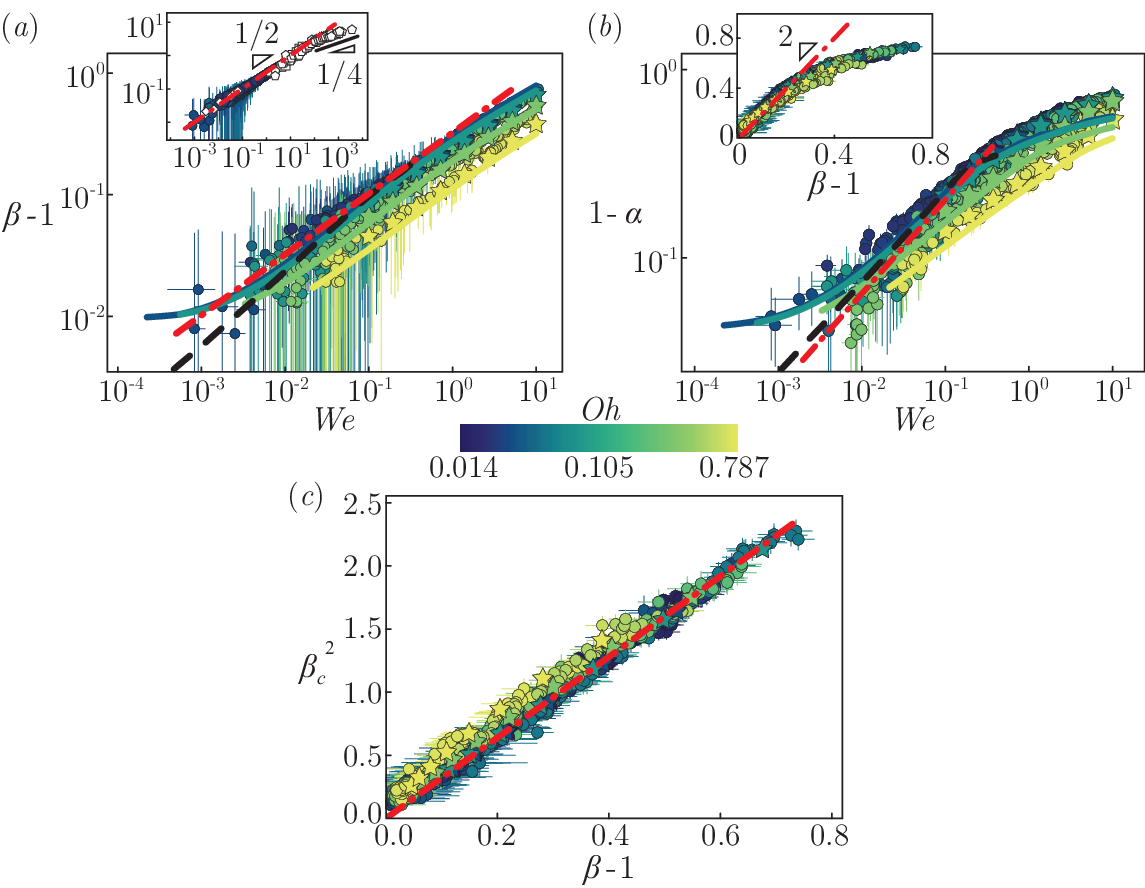}
  \caption{($a$) Maximum equatorial deformation $\beta - 1$ ($\beta = R_{m} / R$) and ($b$) maximum vertical deformation $1-\alpha$ ($\alpha = H_{min} / 2 R $) plotted against Weber number $\We$. Circles represent experiment and stars denote DNS for $\Bo=0.02$. Solid curves show the KM model with $\Bo = 0.02$, and the inertio-capillary limit ($\Oh=0$; $\Bo=0$) is represented by black dashed curves. The red dot-dashed curves are predictions from our simplified model (\S~\ref{subsubsec:energyarguments}). The inset in ($a$) shows our low-viscosity data $\Oh < 0.05$ (colored circles) with prior results (diamond \citep{okumura2003water}, pentagon \citep{clanet2004maximal}, square \citep{biance2006elasticity}, triangle \citep{laan2014maximum}). The red dot-dashed curve is equation (\ref{Beta_We}), and the black solid line with slope 1/4 is used to guide the eye. The inset in ($c$) plots $1-\alpha$ against $\beta - 1$ for all experiments and the red dot-dashed curve is equation (\ref{Alpha_Beta}). ($c$) Dimensionless spreading area $\beta_{c}^{2}$ against $\beta - 1$ for all experiments and DNS results, with equation (\ref{BetaC_Beta}) shown as a  red dot-dashed curve. The color bar maps $\Oh$ to color on a logarithmic scale.}
  \label{fig:dropshape}
\end{figure}

Figure~\ref{fig:dropshape}($a$) shows the maximum equatorial deformation $\beta - 1$ against Weber number $\We$ for all experiments (circles) and DNS (stars). Predictions from the KM model are shown as solid lines for $\Bo=0.02$, with the black dashed curve showing the inertio-capillary limit ($\Oh=0$; $\Bo=0$). The maximum deformation $\beta - 1$ monotonically increases with $\We$ and is well captured by the KM model. The red dot-dashed curve is equation (\ref{Beta_We}), which gives an excellent prediction for our low-viscosity ($\Oh<0.05$) data with no fitting parameters. This agreement persists across four orders of $\We$, as shown more clearly in the inset with our low-viscosity results and data from prior literature \citep{okumura2003water,clanet2004maximal,biance2006elasticity,laan2014maximum}. Beyond $\We \approx 10^{2}$, equation (\ref{Beta_We}) no longer captures the trend of the amalgamated data, which follows a trend more closely resembling $\beta - 1 \sim We^{1/4}$. 

The maximum vertical deformation $1-\alpha$ can also be accurately predicted by our KM model and energy arguments. Figure~\ref{fig:dropshape}($b$) shows $1- \alpha$ against Weber number $\We$ for all experiments (circles) and DNS (stars). The KM model is shown as solid lines for $\Bo=0.02$, and the black dashed line denotes the inertio-capillary limit ($\Oh=0$; $\Bo=0$). The red dot-dashed line is equation (\ref{Alpha_We}). For low $\Oh$ and $\We$, the data follows a power-law trend captured by the KM model and energy argument, but flattens when $\We \gtrapprox 0.5$ due to self-occlusion of the drop surface, a result of our definition for $H$, and because $1-\alpha$ must eventually saturate to 1. The KM model and DNS results capture this feature for all $\Oh$, closely following experimental results. As $\Oh$ increases, $1-\alpha$ decreases and follows a weaker power-law trend. Additionally, self-occlusion is delayed to larger $\We$ or disappears altogether for our largest $\Oh$. The inset in figure~\ref{fig:dropshape}($b$) shows $1-\alpha$ against $\beta-1$ for all experiments and DNS. The data follows the linear trend predicted by equation (\ref{Alpha_Beta}) for $\beta-1 \lessapprox 0.3$. At higher deformations, the data continues to follow a single trend, but necessarily deviates from the energy argument since the deformations are no longer small and $1-\alpha$ must saturate to unity.

The maximum spreading radius $\beta_c$ is a key parameter in determining the heat transfer and total interaction between a drop and substrate in many applications. When $\We > 1$, an impacting drop flattens into a pancake-like shape, such that $\beta_{c} \approx \beta - 1$, making $\beta - 1$ a convenient and practical proxy for $\beta_c$. This is invalid at low $\We$, and necessitates a revised relationship between $\beta_c$ and $\beta$. From our energy argument, equation (\ref{BetaC_Beta}) predicts a linear relationship between the maximum drop deformation and maximum spreading area $\beta_c^{2}$. Figure~\ref{fig:dropshape}($c$) plots $\beta_c^{2}$ against $\beta - 1$ for all experiments (circles) and DNS (stars), with equation (\ref{BetaC_Beta}) overlaid as a red dot-dashed curve. Strikingly, this prediction is accurate across all dimensionless parameters explored, and provides a simple relationship connecting drop spreading and deformation in the low $\We$ regime.

\section{Discussion} \label{sec:discussion}

Previous attempts to quantify rebound behavior for $\We<1$ used superhydrophobic substrates, where contact line friction altered and ultimately suppressed rebound \citep{wang2022successive,thenarianto2023energy}. We overcame this limitation by using a thin viscous film as the substrate, which allowed rebound at as low as $\We=0.0008$. Additionally, we provide the first systematic study of the role of $\Oh$ for $\We<1$ rebounds. Our main finding is a transition to unique trends in the rebound metrics---the coefficient of restitution $\varepsilon$ and contact time $t_c$---at low $\We$. These trends were shown to coincide with a change in the spreading and retraction dynamics, which become symmetric in the inertio-capillary limit, leading to nearly elastic rebound. We complemented these findings with reduced-order models formulated to describe low-$\We$ rebound, and show that they provide accurate predictions for the rebound metrics, drop deformation, and drop spreading.

The coefficient of restitution $\varepsilon$ and contact time $t_c$ have different $\We$-dependencies at low and high $\We$. When $\We \gtrapprox 0.5$, the coefficient of restitution $\varepsilon$ decreased as $\We$ increased, as observed in prior studies \citep{biance2006elasticity,jha2020viscous}. When $\We \lessapprox 0.5$, $\varepsilon$ reached a maximum that depends upon $\Oh$ and $\Bo$, and becomes $\We$-independent in the inertio-capillary limit. A gravity-induced roll-off in $\varepsilon$ leading to $\varepsilon \rightarrow 0$ is the natural route to rebound suppression when $\Bo$ is finite, with $\Oh$ increasing the critical Weber number $\We_c$ below which rebound stops. These results differ from the peak and subsequent roll-off in $\varepsilon$ as $\We$ decreases reported for $\We<1$ rebounds on superhydrophobic substrates \citep{richard2000bouncing,wang2022successive,thenarianto2023energy}. Although these prior experiments were conducted in low $\Oh$ conditions, and in some cases used low $\Bo$ drops, the increased effect of contact line friction at low $\We$ was shown to be the primary mechanism for rebound suppression. Consequently, our results show that submillimeter drops, like those from coughing \citep{dbouk2020coughing} or agricultural sprays \citep{kooij2018determines,makhnenko2021review,chen2022droplet}, may continue bouncing longer than previously thought when the substrate is non-wetting---an important insight into the spread of their contents. The contact time between the drop and substrate is $\We$-dependent at low $\We$ \citep{foote1975water,okumura2003water}, and our results show that it is also $\Oh$-dependent in this regime.  Increasing $\Oh$, causes $t_c$ to increase at a higher rate as $\We$ decreases such that viscosity is most influential on $t_c$ in the low-$\We$ regime. Since $t_c$ is a critical parameter governing drop--substrate interaction in various applications \cite{shiri2017heat,diaz2022charging}, this finding has immediate practical implications, and motivates the study of low $\We$ rebound using drops with shear-dependent viscosity in future work.

In section~\ref{subsec:spreadingandretraction}, we compared the drop shape during spreading and retraction for three values of $\We$ spanning the transition from low $\We$ to high $\We$, and revealed a nearly symmetric spreading and retraction process for low $\Oh$, low $\We$ rebound. As $\We$ increased, the spread time decreased relative to the retraction time, increasing asymmetry between spreading and retraction and aligning with the changes in rebound behavior. \edit{This change corresponds to a loss of energy during rebound when $\We>1$ and was investigated through a careful energy budget analysis by \citet{gilet2012droplets}. Their results showed that when $\We<1$, the inital translational energy was efficiently converted to surface energy during the spreading phase. When $\We>1$, this conversion was inefficient, with an appreciable loss of energy during spreading. This suggests that the time symmetry is broken in the spreading phase at high $\We$. Additionally, \citet{bartolo2005retraction} identified a limit on the retraction rate of a drop that is substrate-independent for low $\Oh$. Although this limit does not explain the onset of asymmetry, which requires considering the energy lost during spreading, it helps rationalize how the spreading-to-retraction time ratio evolves as $\We$ increases when $\We>1$.} \edit{In our experiments, this ratio follows the scaling $t_{s}/t_{r} \sim \We^{-0.26 \pm 0.01}$ in the low $\Oh$ limit. We can rationalize this observation by considering the inertio-capillary limit, where the time needed for a drop to spread $t_{s} \sim R_{c} / V$, where $R_{c} \sim R \, \We^{-1/4}$ from our energy argument (\ref{BetaC_We}), and the maximum retraction rate $t_{r}^{-1} \sim t_{\sigma}^{-1}$ \citep{bartolo2005retraction}. This analysis gives $t_{s} / t_{r} \sim \We^{-1/4}$, in good agreement with our results.} \edit{We therefore interpret the growing asymmetry in rebound dynamics at $\We > 1$ as the result of two effects: a loss of energy during spreading that breaks time symmetry early in the rebound process \citep{gilet2012droplets}, and a limited retraction rate that further skews the rebound timescale \citep{bartolo2005retraction}.  Recent studies have also noted the near-symmetry between spreading and retraction times low $\We$ \citep{lin2024probing}. Our results confirm and extend this observation, by demonstrating that time symmetry holds in the inertio-capillary regime and is progressively lost as either $\We$ or $\Oh$ increases.} 

\edit{Lastly, we examined the drop deformation and spreading in section~\ref{subsec:DropShape}, and showed that the KM model captures drop deformation well across a wide range of $\Oh$, and agrees with a simple energy model in the low $\Oh$ limit, indicating that deformation is primarily in the $l=2$ mode.} The literature on the maximum horizontal deformation is vast and two competing scaling arguments have gained support: one based on momentum conservation $\beta \sim \We^{1/4}$ \citep{clanet2004maximal,bartolo2005retraction,biance2006elasticity,tsai2011microscopic,garcia2020inclined} and the other arising from energy conservation $\beta \sim We^{1/2}$ \citep{chandra1991collision,bennett1993splat,okumura2003water,eggers2010drop,villermaux2011drop,laan2014maximum,lee2016universal}. Despite extensive efforts, the correct scaling remains debated, in part due to most studies using the maximum equatorial radius $\beta = R_{m} /R$ to describe horizontal deformation. However, as discussed in \citep{josserand2016drop}, $\beta$ ignores the initial drop radius and is inadequate to discern scaling arguments at low $\We$. In this work we used the maximum equatorial deformation $\beta-1$, and energy arguments predict $\beta-1 = \left(5/48\right)^{1/2} \, We^{1/2}$, as in \citet{richard2000bouncing}. This prediction is in excellent agreement with our low $\Oh$ data and requires no fitting parameters. Furthermore, we combined our low $\Oh$ data with that from previous studies and validated equation (\ref{Beta_We}) from $\We = 10^{-4}$ to $\We = 10^{2}$, above which the data begins following a trend closer to $ \beta - 1 \sim \We^{1/4}$. As $\Oh$ increases, (\ref{Beta_We}) is less accurate; however, the KM model continues to provide accurate predictions. The KM model also captures the maximum vertical deformation $1-\alpha$ across all $\Oh$, and our energy argument (equation (\ref{Alpha_We})) provides a simple and accurate prediction when $\Oh$ and $\Bo$ effects are negligible. In addition to relating vertical and horizontal deformation, our energy argument allows us to relate maximum deformation and spreading. For example, equation (\ref{BetaC_Beta}) predicts a linear relationship between the horizontal deformation and maximum spread area $\beta_c^2$, which holds for all $\Oh$ and $\Bo$ tested. This relationship provides a useful tool for determining the contact area of a bouncing drop from its deformation, which is critical for applications such as heat transfer \citep{shiri2017heat} or predicting the net charge acquired by a rebounding water drop \citep{diaz2022charging}.

\section{Conclusions} \label{sec:conclusions}

We studied a drop rebounding from a non-wetting, rigid substrate at low $\We$ using experiments, direct numerical simulation, and reduced-order modeling. We \edit{revealed the} rebound behavior \edit{of drops} at low $\We$ for a range of $Oh$, \edit{showed how increasing $Oh$ or $We$ breaks the symmetry between the spreading and retraction phases}, and elucidated the critical role of $\Oh$ and $\Bo$ in suppressing rebound. The maximum drop deformation and spreading were connected through energy arguments, and closely followed the predictions of the KM model irrespective of $\Oh$. These findings reveal new physical insight into drop rebound and provide practical tools for predicting drop deformation on non-wetting substrates.  

The interplay between our theoretical approaches, the KM model and the dedicated direct numerical simulation setup, enabled us to cross-validate these tools and establish the range of validity of the reduced-order model with a significant level of confidence, with the experimental dataset acting as foundation where comparisons were possible. Thus, we were suitably equipped to use the more efficient methodology in the relevant section of the parameter space and relevant distinguished limits thereof, while also being able to rely on DNS for larger values of $\We$ where the model assumptions no longer hold. \edit{In general, DNS is preferable for $\We > 1$, while the reduced-order model offers high-accuracy predictions with substantially lower computational cost for $\We < 1$, but continues to yield reasonable estimates up to $\We \approx 10$.} Furthermore, our work also presents the first application of the kinematic match method to the case of a deformable impactor rebounding off a rigid substrate. The quality of the resulting predictions highlights its versatility and naturally invites further applications for it; e.g., its use in the case of a collision of two deformable objects, and in combination with moving meshes and variational methods. 

When contact line formation is avoided, the coefficient of restitution $\varepsilon$ is $\We$-independent at low $\We$ in the inertio-capillary limit. Finite $\Bo$ effects suppress bouncing while increasing $\Oh$ decreases $\varepsilon$ and increases the critical Weber number $\We_{c}$ where rebound ceases. In contrast, the contact time $t_{c}$ increases as $\We$ decreases at a rate that increases with $\Oh$. The shift in how $\varepsilon$ and $t_{c}$ change with $\We$ in the low $We$ and high $We$ regimes coincides with altered spreading and retraction dynamics. In particular, the nearly elastic rebound observed in the inertio-capillary limit have a nearly-symmetric spreading and retraction phase, but increasing $We$ above unity and increasing $Oh$ decreases $\varepsilon$ and increases asymmetry in the spreading and retraction process. Using energy arguments, we derive simple equations that accurately predict the maximum vertical and horizontal deformation, and the maximum spreading for low $\Oh$ impacts with no fitting parameters. The prediction for the maximum horizontal deformation (\ref{Beta_We}) reveals that the horizontal deformation of an impacting drops scales as $\We^{1/2}$, which when compared with experimental data holds up to $\We\approx 10^{2}$. Additionally, this simple model accurately relates $\beta-1$ to the maximum vertical deformation $1-\alpha$ via equation (\ref{Alpha_Beta}) and maximum contact area $\beta_{c}^{2}$ via equation (\ref{BetaC_Beta}). For all $\Oh$, the kinematic match model provides excellent predictions for the maximum spreading and deformation for low $\We$ rebound where direct numerical simulation struggles due to numerical stiffness and parasitic capillary currents.

Although we focus here on the normal impact of a drop on a rigid surface, the ubiquity of drop impacts make our findings relevant to many situations. For example, low-$\We$ rebound governs the final stage of drop dispersal in many applications involving low-$\Bo$ drops, such as agricultural sprays \citep{kooij2018determines,makhnenko2021review,chen2022droplet} and cough ejecta \citep{dbouk2020coughing,bourouiba2021fluid}, where understanding the spread of the drops contents is essential. \edit{In both examples, the drops typically impact non-wetting surfaces, as plant leaves are often naturally superhydrophobic \citep{barthlott1997purity}, while medical surfaces are frequently rendered superhydrophobic to reduce microbial adhesion and biofluid contamination \citep{falde2016superhydrophobic}.} Future work is needed to extend our results to more complex practical and natural settings, which often involve non-Newtonian drops and complaint substrates that have a pronounced effect on drop impact at high $\We$ \citep{bartolo2007dynamics,basso2020splashing} but remain unexplored for drop rebound at low Weber number. Furthermore, our focus herein has been strictly on the liquid dynamics whereas the coupled air layer dynamics are also a source of rich physics, and have become of increasing interest to modelers \cite{Phillips_Milewski_2024,sprittles2024gas}.

\section*{Supplementary movies.} Supplementary movies are available at \\ \url{https://github.com/harrislab-brown/LowWeberDropRebound}. 

\section*{Supplementary material.}
The kinematic match code, supplemental movies, and data used in each figure are available at \url{https://github.com/harrislab-brown/LowWeberDropRebound}. The direct numerical simulation code infrastructure, including pre- and post-processing functionality and a short descriptive tutorial, can be found at \\ \url{https://github.com/rcsc-group/BouncingDropletsSolidSurface}.

\section*{Funding.}
The authors gratefully acknowledge the financial support of the National Science Foundation (NSF CBET-2123371) and the Engineering and Physical Sciences Research Council (EPSRC EP/W016036/1).  C.T.G. gratefully acknowledges the Hope Street Fellowship for financial support. 

\section*{Declaration of Interests.}
The authors report no conflict of interest.

\section*{Data accessibility.}
All data reported in this study is available at \\ \url{https://github.com/harrislab-brown/LowWeberDropRebound}. 

\appendix

\section{Defining contact}\label{appA}

\begin{figure}
  \centering
  \includegraphics[width=135mm]{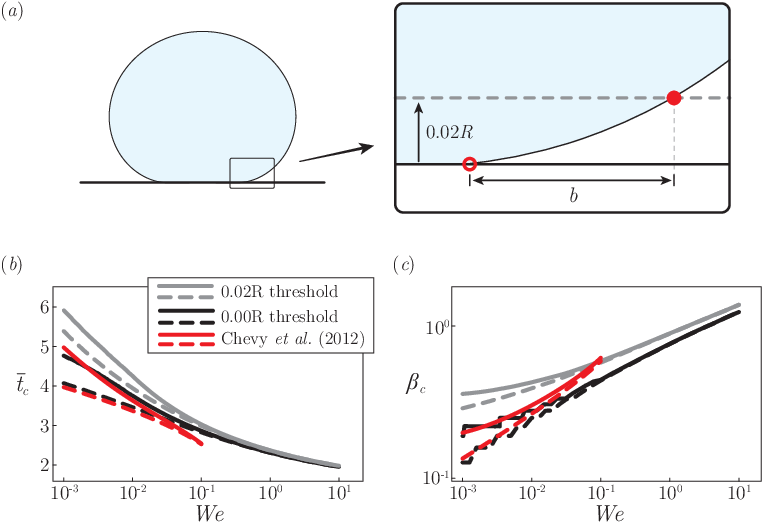}
  \caption{($a$) Magnified view near a drop's contact point during impact. The hollow circle indicates the contact point with the substrate, while the filled circle marks the contact point along a hypothetical substrate at a height corresponding to the experimental spatial resolution $z=0.02R$. This threshold increases the contact time and expands the contact radius $r_c$ by $b$. ($b$) Dimensionless contact time $\bar{t}_{c} = t_{c} / t_{\sigma}$ and ($c$) dimensionless maximum contact radius $\beta_{c} = R_{c} / R$ against Weber number for the KM model evaluated at the hypothetical substrate $0.02 R$ (gray) and true substrate $0.00 R$ (black) for $\Oh = 0.03$. Red curves represent predictions from the model of \citet{chevy2012liquid}. Solid lines show predictions for $\Bo=0.019$, and dashed lines show results obtained for $\Bo=0$.}
\label{fig:threshold}
\end{figure}

Determining the substrate location is crucial for evaluating drop–substrate interactions. However, this is typically constrained by the spatial resolution of the measurement technique. In this study, the presence of an intervening air layer further complicates identifying the substrate, as it prevents direct contact with the drop. In our experiments the spatial resolution related to the pixel width in our images was approximately $0.02R$, which can lead to discrepancies when compared with KM model and simulation results if not considered. Figure~\ref{fig:threshold}($a$) shows a drop impacting a substrate with a magnified view of the contact boundary. Increasing the substrate location by the spatial resolution of $0.02R$ (gray dashed line) shifts the true contact point (open circle) to the perceived contact point (solid circle), expanding the contact radius by an amount $b$ and necessitating careful criteria for faithfully comparing experimental and modeling results.  The non-wetting nature of the surface makes this process particularly sensitive due to the tangency with the substrate.

To provide the most accurate predictions between the KM model, DNS, and experiments, the former two measure all relevant variables using  a hypothetical substrate located $0.02R$ above the true substrate. This criteria also allows DNS measurements to be made above the intervening air layer, which would otherwise need to be accounted for during post-processing. The parameters most affected by this criterion are the contact time, $t_c$, and contact radius, $r_c$, which both increase as a result. To show the magnitude of this effect, we plot the dimensionless contact time $t_c / t_{\sigma}$ against $\We$ for the KM model with (gray lines) and without (black lines) the $0.02R$ criteria in figure~\ref{fig:threshold}($b$), where solid lines represent $\Bo = 0.019$ and dashed lines are for $\Bo = 0$. The influence of the threshold increases as $\We$ decreases. Figure~\ref{fig:threshold}($c$) shows the KM model for the dimensionless maximum contact radius $\beta_{c} = R_{c} / R$ versus $\We$ for the same threshold values. Again, the threshold effect increases as $\We$ decreases; however, unlike contact time, noticeable differences in the predictions persist even for $\We>1$. 

By employing a discretized angular mesh for the polar coordinate, only a finite number of contact radii are numerically allowed. As a consequence, the $0.00R$ criterion predicts piecewise constant functions for the contact radius $r_c$, while the $0.02R$ criterion smooths this discretized behavior and results in smoother curves, as seen in figure~\ref{fig:threshold}($c$). 

We also compare these predictions to results obtained using the quasi-static model of \citet{chevy2012liquid}.  Their work predicts contact time $t_c$, but additional features about the shape, such as the contact radius $r_c$, are also extractable by reproducing their calculations.  In figures~\ref{fig:threshold}($b,c$), the red curves show predictions form \citet{chevy2012liquid}'s model with the solid line for $\Bo=0.019$ and dashed line for $\Bo=0$. While their modeling approach differs, their low-$\We$ predictions align very well with ours, but the KM model is able to make predictions above $\We \approx 10^{-1}$ while also enforcing the natural geometric constraints in the contact region at all points.

\section{Numerical approximations} \label{appB}

\subsection{The spectral method} \label{subsubsec:thespectralmethod}

We define $A_l$, $U_l$, $B_l$, $\aproxsol{h}$ and $\aproxsol{v}$ as approximations of 
$\mathscr{A}_l$, $\mathscr{U}_l$, $\mathscr{B}_l$, $h$ and $v$, respectively. These represent the solutions to the system equations obtained by truncating the expansions of the surface shape, velocity and pressure distribution in spherical harmonics up to the L$^{th}$ mode. That is
\begin{equation}\label{eqn:nd_dot_A_l_v01}
    \dot{A}_l = U_l, \qquad \text{ for } l = 2,\ldots,L;
\end{equation}
\begin{equation}\label{eqn:nd_dot_U_l_v01}
    \dot{U}_l 
    =
    -l \left(l+2\right)\left(l-1\right)A_l -2(2l+1)(l-1)\,  \Oh \, {U}_l
    -l \, B_l, \qquad \text{ for } l = 2,\ldots, L;
\end{equation}
\begin{equation}\label{eqn:nd_dot_h_v}
    \dot{\aproxsol{h}} = \aproxsol{v};
\end{equation}
\begin{equation} \label{eqn:nd_dot_v}
    \dot{\aproxsol{v}} = - \Bo + B_1;
\end{equation}
which are subject to 
\begin{equation}
    A_l(t=0) = 0, \qquad \text{ for } l = 2,\ldots, L;
\end{equation}
\begin{equation}
    U_l(t=0) = 0, \qquad \text{ for } l = 2,\ldots, L;
\end{equation}
\begin{equation}
    \aproxsol{h}(t=0) = 1;
\end{equation}
\begin{equation}
    \aproxsol{v}(t=0) = -\sqrt{\We};
\end{equation}
\begin{equation}
    \sum\limits_{l=2}^{L} {A_lP_l\left(\cos(\theta)\right)} = \frac{\aproxsol{h}}{\cos(\theta)}-1, \qquad \text{ for } \theta \leq \theta_c;
\end{equation}
\begin{equation}
    \sum\limits_{l=2}^L {A_lP_l(\cos(\theta))} > \frac{\aproxsol{h}}{\cos(\theta)}-1, \qquad \, \text{ for } \theta > \theta_c;
\end{equation}
\begin{equation}
    \sum\limits_{l=0}^{L} {B_lP_l(\cos(\theta))} = 0, \qquad \hspace{15 mm} \text{ for } \theta > \theta_c;
\end{equation}
\begin{equation}
    \frac{\partial}{\partial \theta}\left[\left(1+\sum\limits_{l=2}^L {A_lP_l(\cos(\theta))}\right)\cos(\theta)\right]_{\theta=\theta_c} = 0.
\end{equation}

For the present work, calculations were performed using $L=90$ spherical modes. Convergence was verified by testing resolutions of 120, confirming that all quantities of interest varied by less than 10\% across the full dimensionless parameter range—representing the worst-case deviation—with an average relative error below 5\%.

\subsection{Finite differences in time} \label{subsubsec:finitedifferencesintime}

We approximate the time derivatives by defining
$A_l^k \approx A_l(t=t^k) \text{, } U_l^k \approx U_l(t=t^k)\text{, } B_l^k \approx B_l(t=t^k) \text{, } h^k \approx \aproxsol{h}(t=t^k) \text{, } v^k \approx \aproxsol{v}(t=t^k)$ and applying the implicit Euler method to equation (\ref{eqn:nd_dot_A_l_v01}),
\begin{equation}\label{eqn:IE_dot_A_l}
    A^{k+1}_l -\delta_t^k U^{k+1}_l = A^k_l \qquad \text{ for } l = 2, \ldots L,
\end{equation}
where $\delta^k_t := t^{k+1}-t^k$.  Similarly, from equation (\ref{eqn:nd_dot_U_l_v01}), we have

\begin{equation}\label{eqn:Euler_mode_speed}
\begin{aligned}
    \delta^k_t \, l \, (l+2)(l-1) \, A^{k+1}_l + \left[1 + \delta^k_t \, 2(2l+1)(l-1) \, \Oh \right] \,{U}_l^{k+1} + \delta^k_t \, l \, B_l^{k+1} =U^k_l,& \\ \text{ for } l = 2, \ldots L,&
\end{aligned}
\end{equation}
from equation (\ref{eqn:nd_dot_h_v}), we have
\begin{equation}
    - \delta_t^k \, v^{k+1} + h^{k+1} = h^{k},
\end{equation}
and from equation 
(\ref{eqn:nd_dot_v}), we have
\begin{equation}\label{eqn:Acc_CM}
    -\delta_t^k \, B_1^{k+1} + v^{k+1} = v^k -  \delta_t^k \, \Bo;
\end{equation}
which are subject to
\begin{equation}\label{eqn:km_contact_amplitudes_finite_modes}
    \sum\limits_{l=2}^{L} {A_l^{k+1}P_l\left(\cos(\theta)\right)} - \frac{h^{k+1}}{\cos(\theta)} = -1,
    \qquad \quad  \hspace{1 mm} \text{ for } \theta \leq \theta_c;
\end{equation}
\begin{equation}\label{eqn:km_non-contact_amplitudes_finite_modes}
    \cos(\theta)\left(1+\sum\limits_{l=2}^{L} {A_l^{k+1}P_l\left(\cos(\theta)\right)}\right) > h^{k+1}, \qquad \text{ for } \theta > \theta_c;
\end{equation}
\begin{equation}\label{eqn:Press_zero}
    \sum\limits_{l=0}^{L} {B_l^{k+1}P_l(\cos(\theta))} = 0, \qquad \hspace{22 mm} \text{ for } \theta > \theta_c;
\end{equation}
\begin{equation}\label{eqn:km_contact_derivative_finite_modes}
    \frac{\partial}{\partial \theta}\left[\left(1+\sum\limits_{l=2}^L {A_l^{k+1}P_l(\cos(\theta))}\right)\cos(\theta)\right]_{\theta=\theta_c} = 0.
\end{equation}

\subsubsection{Angular discretization} \label{subsubsec:angulardiscretisation}

The spectral mesh is determined by the maximum order of harmonic modes $L$ in our truncated expansion, using Legendre polynomials of degree $0$ to $L$. The angular mesh is naturally defined as the union of ${\theta = 0}$ (the south pole) and the zeros of $P_L(\cos(\theta))$, which is canonical for the Legendre polynomial basis \citep{boyd2000chebyshev}. A schematic of this angular mesh is shown in figure~\ref{fig:model_graphic}($c$).

The candidate integer $q$ defines the number mesh points from the south pole to the boundary of the contact area, where the pressure is allowed to be non-zero. The boundary angle of the contact area is taken as the midpoint between $\theta_q$ and $\theta_{q+1}$. This results in the discrete version of the constraints imposed by \eqref{eqn:km_contact_amplitudes_finite_modes}-\eqref{eqn:km_contact_derivative_finite_modes}
\begin{equation}\label{eqn:match_contact}
    \sum\limits_{l=2}^{L} {A_l^{k+1, q}P_l\left(\cos(\theta_i)\right)} - \frac{h^{k+1, q}}{\cos(\theta_i)} = -1,
    \qquad \hspace{4.6 mm} \text{ for } i \leq q; 
\end{equation}
\begin{equation}\label{eqn:no_contact}
    \cos(\theta_i)\left(1+\sum\limits_{l=2}^{L} {A_l^{k+1, q}P_l\left(\cos(\theta_i)\right)}\right) > h^{k+1, q},
    \qquad \text{ for } i > q;
\end{equation}
\begin{equation}\label{eqn:Press_zero}
    \sum\limits_{l=0}^{L} {B_l^{k+1, q}P_l(\cos(\theta_i))} = 0, \qquad \hspace{23.2 mm} \text{ for } i > q; 
\end{equation}
and
\begin{equation}\label{eqn:tan_cond_q}
    \frac{\partial}{\partial \theta}\left[\left(1+\sum\limits_{l=2}^L {A_l^{k+1,q}P_l(\cos(\theta))}\right)\cos(\theta)\right]_{\left(\theta_q+\theta_{q+1}\right)/2} = 0. 
\end{equation}
Here, $A_l^{k+1,q}$ and $B_l^{k+1,q}$ are the amplitudes of the surface deformation and pressure distribution, respectively, obtained by solving the discrete forms of equations \eqref{eqn:match_contact}, \eqref{eqn:Press_zero}, \eqref{eqn:IE_dot_A_l}-\eqref{eqn:Acc_CM}, assuming exactly $q$ mesh points in the contact area:
\begin{equation}\label{eqn:IE_dot_A_l}
    A^{k+1, q}_l -\delta_t^k \, U^{k+1, q}_l = A^k_l, \qquad \text{ for } l = 2, \ldots L,
\end{equation}
\begin{equation}\label{eqn:Euler_mode_speed}
\begin{aligned}
    \delta^k_t\, l \, (l+2)(l-1) \, A^{k+1, q}_l + \left[1 + \delta^k_t \, 2(2l+1)(l-1) \, \Oh \, \right] \, {U}_l^{k+1, q} + \delta^k_t \, l \, B_l^{k+1, q} =U^k_l,& \\ \text{ for } l = 2, \ldots L,&
\end{aligned}
\end{equation}
\begin{equation}\label{eqn:acc_cm_q}
    - \delta_t^k \, v^{k+1, q} + h^{k+1, q} = h^{k},
\end{equation}
and
\begin{equation}\label{eqn:Acc_CM}
    -\delta_t^k \, B_1^{k+1,q} + v^{k+1, q} = v^k -  \delta_t^k \, \Bo,
\end{equation}
where $q$ in the superscript denotes solutions calculated assuming $q$ contact points.

\subsection{A discrete optimization problem} \label{subsubsec:adiscreteoptimizationproblem}

For a given number of contact mesh points $q$, we can form a closed, linear system of equations with \eqref{eqn:Press_zero}, \eqref{eqn:match_contact} and \eqref{eqn:IE_dot_A_l}-\eqref{eqn:Acc_CM}. We use equation \eqref{eqn:no_contact} to define 
\begin{equation}\label{eqn:cases_impenetrability}
    e^1_q
    =
    \begin{cases}
        0,\  &\cos(\theta_i)\left(1+\sum\limits_{l=2}^{L} {A_l^{k+1, q}P_l\left(\cos(\theta_i)\right)}\right) > h^{k+1, q},
    \  \text{ for all } i > q; \\
        \infty,\  &\text{otherwise}.
    \end{cases}
\end{equation}
This discrete version of condition \eqref{eqn:no_contact} ensures the drop does not touch or penetrate the substrate outside the contact area.

For each $q$, the values of $A_l^{k+1,q}$  are substituted into the second-order approximation of equation \eqref{eqn:tan_cond_q}, and the absolute value of the resulting residual is computed as
\begin{equation}\label{eqn:discrete_tangential_error}
    e^2_q = \left|
    \left[1+
    \sum\limits_{l=2}^{L} {A_l^{k+1, q}P_l\left(\cos(\theta_q)\right)}
    \right]\cos\left(\theta_q\right)
    -
    \left[1+
    \sum\limits_{l=2}^{L} {A_l^{k+1, q}P_l\left(\cos(\theta_{q+1})\right)}
    \right]\cos\left(\theta_{q+1}\right)
    \right|.
\end{equation}
Finally, we define
\begin{equation}
    e_q = \max \left\{e^1_q,e^2_q\right\}.
\end{equation}
The value of $q$ that minimises $e_q$ defines the contact mesh points at time $k+1$; that is,
\begin{equation}\label{eqn:Objective_func_discrete}
    m^{k+1} 
    \coloneqq 
    \text{argmin}_q 
    e_q,
\end{equation}
and, consequently,

\begin{align}
    & A_l^{k+1} \coloneqq A_l^{k+1,m^{k+1}}, \ h^{k+1} \coloneqq h^{k+1,m^{k+1}}, \
    v^{k+1} \coloneqq v^{k+1,m^{k+1}}, \\
    & B_l^{k+1} \coloneqq B_l^{k+1,m^{k+1}}, 
    \text{ and }\ \ 
    U_l^{k+1} \coloneqq U_l^{k+1,m^{k+1}}.
\end{align}

\subsection{The system matrices} \label{subsubsec:thesystemmatrices}

The discretized problem involves solving a linear system of equations at each sampling time for each candidate contact area radius, parameterized by $q$. These systems can be expressed in matrix form as:
\begin{equation}\label{eqn:Sys_eqn_k+1_q}
M^{k+1,q} \, x^{k+1,q} = b^{k+1,q},
\end{equation}
where
\begin{equation}
x^{k+1,q}
=
\left[ A_2^{k+1,q},
       \ldots,
       A_L^{k+1,q},
       U_2^{k+1,q},
       \ldots,
       U_L^{k+1,q},
       B_0^{k+1,q},
       \ldots,
       B_L^{k+1,q},
       h^{k+1,q},
       v^{k+1,q}
\right]^T ,
\end{equation}
\begin{equation}
    b^{k+1,q} = \left[A_2^k,\ldots,A_L^k,U_2^k,\ldots,U_L^k,\underbrace{-1,\ldots,-1}_{q \text{ times}},\underbrace{0,\ldots,0}_{L+1-q \text{ times}},h^k,v^k-\delta_t^{k} \, \Bo \right]^T,
\end{equation}
and
\begin{equation}
    {\bf M}^{k+1,q} =
    \left[
    \begin{array}{ccccc}
        I_{L-1} & -\delta_t^kI_{L-1} & 0& 0 & 0 \\
        \delta_t^k C & D & \delta_t^k E & 0 & 0 \\
        F^q & 0 & 0 & G^q & 0 \\
        0 & 0 & H^q & 0 & 0 \\
        0 & 0 & 0 & 1 & -\delta_t^k \\
        0 & 0 & \delta_t^k K & 0 & 1
    \end{array}
    \right].
\end{equation}
Here, $I_{m}$ is the identity matrix of size $m$, and from equation \eqref{eqn:Euler_mode_speed} we have
\begin{equation}
    C_{i,j} = j \, (j+1)(j+3) \, \delta_{i,j},
\end{equation}
\begin{equation}
    D_{i,j} = \left(1 +  \delta^k_t \, 2 \, \Oh \,j \, (2j+3)\right) \, \delta_{i,j}.
\end{equation}
For all $i, j = 1, 2, \dots, L-1$, where $\delta_{i, j}$ is the Kronecker delta. Also, 
\begin{equation}
    E = \left[
    \begin{array}{cc|ccccc}
       0      & 0      &  2      & 0      &\ldots &\ldots    &0\\
       0      & 0      &  0      & 3      &0      &\ldots    &0\\
       \vdots & \vdots &  \vdots & 0      &\ddots &          &\vdots\\
       \vdots & \vdots &  \vdots & \vdots &       &(L-1)     &0\\
       0      & 0      &  0      & 0      &0      &\ldots    &L\\
    \end{array}
    \right],
\end{equation}
where the first two zero columns represent modes $0$ and $1$, absent in equation \eqref{eqn:Euler_mode_speed}. 

Similarly, from equation (\ref{eqn:match_contact}), 
\begin{equation}
    F_{i,j}^q = P_{j+1}(\cos(\theta_{i-1})), 
    \qquad \text{ for } i= 1,2,\ldots,q; 
    \text{ and } j = 1,2,\ldots,L-1;
\end{equation} 
and
\begin{equation}
    G_i^q = \left [
    \frac{-1}{\cos(\theta_i)}    
    \right ], \qquad \text{ for } i= 1,2,\ldots,q.
\end{equation}
From equation (\ref{eqn:Press_zero}) we have
\begin{equation}
    H_{i,j}^q = P_{j-1}(\cos(\theta_{i + q-1})), 
    \qquad \text{ for } i= 1, 2,\ldots,L- q +1; \text{ and } j = 1,2,\ldots,L+1;
\end{equation} 
and from (\ref{eqn:Acc_CM}) we obtain
\begin{equation}
    K = \left[\underbrace{
    \begin{array}{ccccc}
       0 & -1 & 0 & \ldots & 0
    \end{array}}_{L+1 \text{ elements}}
    \right].
\end{equation}

In summary, the discrete problem is reduced to finding the $q$ that satisfies equation (\ref{eqn:Objective_func_discrete}). Instead of solving system (\ref{eqn:Sys_eqn_k+1_q}) for every $q$, we solve it for $q = m^k-2, m^k-1, m^k, m^k+1, m^k+2, $. This approach significantly reduces computational costs without decreasing accuracy, enabling typical runs to complete within minutes.

\section{Computational implementation}\label{appC}

\subsection{The adaptive time step} \label{subsubsec:theadaptivetimestep}

We set a maximum dimensionless step size of $2\pi/\left(S_L \left( L(L+2)(L+1) \right)^{1/2} \right)$, where $S_L = 16$, chosen as a fraction $1/S_L$ of the oscillation period of the fastest harmonic mode. The adaptive time step refines the initial time mesh by adding discrete times as needed. The step size is dynamically adjusted to ensure the contact area evolution is resolved with the highest accuracy permitted by the angular mesh. Specifically, if the model predicts that the number of contact mesh points $m^{k}$ changes by more than one in a single step, the step is rejected, halved, and candidate solutions are recalculated. This approach relies on the assumption that the contact area evolves continuously during impact.

\begin{figure}
  \centering
  \includegraphics[width=130mm]{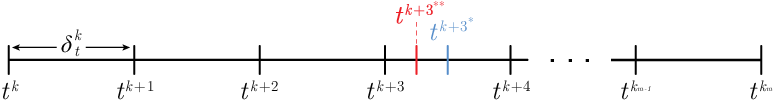}
  \caption{Time line of regularly spaced intervals (black) of spacing $\delta_{t}^{k}$. Blue and red lines represent the time after one and two refinement steps between $t^{k+3}$ and $t^{k+4}$, respectively, where $\delta_{t}^{k}$ is halved at each.}
\label{fig:adaptive_time_step}
\end{figure}

The code iteratively refines the time step as needed to ensure the motion of the contact area's boundary is resolved to the accuracy of the angular mesh. Figure~\ref{fig:adaptive_time_step} illustrates two consecutive time step refinements between $t^{k+3}$ and $t^{k+4}$, following the rejection of the initial calculation $t^{k+4}$. This results in the insertion of $t^{k+3^*}$, which, if also rejected, leads to the insertion of $t^{k+3^{**}}$. While not shown in the figure, the discrete time sequence is renumbered to include these refinements.

\subsection{Finding local minima of the error} \label{subsubsec:findinglocalminimaoftheerror}

\begin{figure}
  \centering
  \includegraphics[width=135mm]{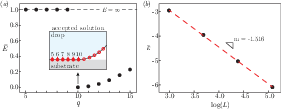}
  \caption{Normalized error $\bar{E} = 2 \, \textrm{tan}^{-1}(E) / \pi$ versus candidate contact point $q$ for a drop with 10 contact points. An infinite error $E = \infty$ (dashed line) is assigned for $q < 10$ using equation \eqref{eqn:cases_impenetrability}. For $q \geq 10$, the error is calculated using \eqref{eqn:discrete_tangential_error} and increases with $q$, showing a minimum at $q=10$. The inset shows a schematic of the accepted solution. ($b$) Relative change in the coefficient of restitution $\tilde{\varepsilon} = \textrm{log} \left( |\varepsilon_L-\varepsilon_{L/2}|/\varepsilon_{L/2} \right)$  against number of modes log($L$). The dashed line is a power law fit to the data with slope m = -1.516.}
\label{fig:error_convergence}
\end{figure}

Assuming the contact area boundary moves continuously, we can reduce the computational cost of solving the discrete optimization problem in equation (\ref{eqn:Objective_func_discrete}), which would otherwise require evaluating all possible $q$ from $0$ to $L$. The error in the angle between the drop and substrate (the quantity to minimize) exhibits clear monotonic behavior around the minimum. Figure~\ref{fig:error_convergence}($a$) shows the normalized error $\bar{E} = 2 \, \textrm{tan}^{-1}(E) / \pi$ plotted against $q$ for a case where the solution indicates that $q=10$ contact points is the optimal solution, i.e., $m^{k+1}=10$. Our normalization of the error allows us to display both finite and infinite errors in the same plot, with $\bar{E}=1$ for $E = \infty$. 

These results indicate that to determine $m^{k+1}$, the number of contact points at time $t^{k+1}$, it is sufficient to test a subset of five values: $q = m^{k}-2$ to $q = m^{k}+2$. We accept only a minimum of $e_q$ within one mesh point of $m^k$, and these five points provide enough information to identify any local minimum among the three acceptable points. If the minimum occurs at $q = m^{k}-2$ or $q = m^{k}+2$, the solution is rejected, and the time step is halved; otherwise, the solution is accepted.

In practice, we first check if $q = m^k-1$ is a local minimum (unless $m^k=0$). To do this, we compute the errors for $q = m^{k}-2$ to $q = m^{k}$. If $q = m^{k}-1$ is the minimum, we accept it as the correct solution. If $q = m^{k}-2$ has the least error, the solution is rejected and the time step is refined. Otherwise, if $q = m^{k}$ has the least error, we check $q = m^{k}+1$. If $q = m^{k}$ is confirmed as a local minimum, the solution is accepted. If not, we compute the error for $q = m^{k}+2$ and either accept $q = m^{k}+1$ as the solution (if it is a local minimum) or reject it and refine the time step before recalculating. 

\begin{figure}
  \centering
  \includegraphics[width=135mm]{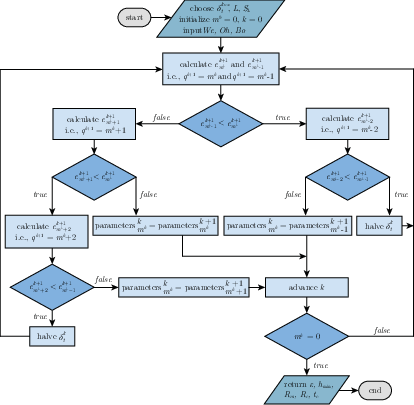}
  \caption{Flowchart of computational implementation. The return variables are the coefficient of restitution $\varepsilon$, minimum height of the center of mass $h_{\text{min}}$, maximum equatorial radius $R_{\text{m}}$, maximum spreading radius $R_c$ and contact time $t_c$.}
\label{fig:code_flowchart}
\end{figure}

We verify whether any $q$ causes the drop interface to intersect the substrate outside the region where pressure is allowed to be nonzero. This check is performed for all points in the angular mesh with $\theta_q<\theta_i<\pi/2$. If an intersection occurs, the error in the discrete optimization is set to $\infty$. To confirm convergence in our results, we plot the relative change in the coefficient of restitution $\tilde{\varepsilon} = \textrm{log} \left( |\varepsilon_L-\varepsilon_{L/2}|/\varepsilon_{L/2} \right)$ for a typical rebound against number of modes log($L$). A power law trend with slope $\textrm{m} = - 1.516$ describes the convergence our results and motivates our choice of $L=90$ throughout this work. A flowchart for this procedure is presented in figure~\ref{fig:code_flowchart}.

\bibliographystyle{jfm}
\bibliography{DropRebound_JFM}

\end{document}